\documentclass[preprint,letterpaper,12pt]{aastex}
\slugcomment{{\sc Accepted for publication in The Astrophysical Journal Supplement:} 2014 June 18}

\usepackage{amsmath}
\usepackage[pdfborder={0 0 1}]{hyperref}
\usepackage{natbib}

\newcommand{\afm}{\altaffilmark}

\newcommand{\madcows}{MaDCoWS}
\newcommand{\spitzer}{\emph{Spitzer}}

\begin{document}

\title{The Massive and Distant Clusters of WISE Survey II: Initial Spectroscopic Confirmation of $z \sim 1$ Galaxy Clusters Selected from 10,000 Square Degrees }
\author{%
  S. A. Stanford\afm{1}$^,$\afm{2}, 
  Anthony H. Gonzalez\afm{3}, 
  Mark Brodwin\afm{4}, 
  Daniel P. Gettings\afm{3}, 
  Peter R. M. Eisenhardt\afm{5}, 
  Daniel Stern\afm{5}, and 
  Dominika Wylezalek\afm{6}
}

\altaffiltext{1}{Department of Physics, University of California, One Shields Avenue, Davis, CA 95616}
\altaffiltext{2}{Institute of Geophysics and Planetary Physics, Lawrence Livermore National Laboratory, Livermore, CA 94550}
\altaffiltext{3}{Department of Astronomy, University of Florida, 211 Bryant Space Center, Gainesville, FL 32611}
\altaffiltext{4}{Department of Physics and Astronomy, University of Missouri, 5110 Rockhill Road, Kansas City, MO 64110}
\altaffiltext{5}{Jet Propulsion Laboratory, California Institute of Technology, 4800 Oak Grove, Pasadena, CA 91109}
\altaffiltext{6}{European Southern Observatory, Karl-Schwarzschildstr.2, D-85748, Garching bei Munchen, Germany}

\begin{abstract}
We present optical and infrared imaging and optical spectroscopy of galaxy clusters which were identified as part of an all-sky search for high-redshift galaxy clusters, the Massive and Distant Clusters of WISE Survey (MaDCoWS).   The initial phase of MaDCoWS combined infrared data from the all-sky data release of the Wide-field Infrared Survey Explorer (WISE) with optical data from the Sloan Digital Sky Survey (SDSS) to select probable $z \sim 1$ clusters of galaxies over an area of 10,000 deg$^2$.  Our spectroscopy confirms 19  new clusters at $0.7 < z < 1.3$, half of which are at $z > 1$, demonstrating the viability of using WISE to identify high-redshift galaxy clusters.  The next phase of MaDCoWS will use the greater depth of the AllWISE data release to identify even higher redshift cluster candidates.

\end{abstract}


\section{Introduction}
\label{intro}

Studies of massive clusters of galaxies have provided the first evidence for the existence of dark matter \citep{zwicky1937},  
showed the importance of environment in galaxy evolution \citep{dressler1980}, 
and currently can provide competitive constraints on cosmological parameters \citep{allen2011,benson2013}.   
Much of the power of such studies derives primarily from the extreme cluster masses ($M>10^{14}$~M$_{\sun}$).  The largest area surveys can identify the rarest, most massive clusters.  The {\it ROSAT} All-Sky Survey resulted in catalogs of very massive  galaxy clusters at low to moderate redshifts \citep[i.e., the BCS at $z<0.3$ and MACS at $z<0.7$]{ebeling1998,ebeling2001}.   More recently, the {\it Planck} mission has provided an all-sky catalog of very massive galaxy clusters at $z<1$ \citep{planck2013}, while the South Pole Telescope (SPT) and Atacama Cosmology Telescope (ACT) detect high-mass clusters reaching even above $z>1$ over several thousand square degrees \citep{williamson_2011,marriage2011,reichardt_2012}.  
However, until recently no surveys were capable of identifying massive clusters at $z\ga1$ over the full extragalactic sky.  

The primary means of identifying clusters of galaxies include optical/infrared searches for galaxy overdensities, X-ray searches that detect the hot gas of the intracluster medium, and mm-wave searches that rely on the Sunyaev-Zel'dovich (SZ) effect.  The latter  technique is nearly mass-limited due to the weak redshift dependence of the detection signal, which makes SZ surveys ideal for cosmological tests based upon evolution of the cluster mass function.  Similar to the SZ, surveys at X-ray wavelengths have the advantage of being able to estimate cluster masses directly from their data (if the cluster redshifts are known).  However, the X-ray signal drops dramatically with redshift, and X-ray surveys so far have not been able to cover very large areas to the depth necessary to find large numbers of very massive clusters at high redshifts.  In the future, {\it eROSITA} \citep{predehl2010} should provide an upgrade to the prospects of large-area X-ray cluster surveys.   

Using infrared imaging, cluster searches that detect galaxy overdensities provide the greatest reach 
in redshift and mass sensitivity at high-redshift, extending to $z \sim 2$ and below $M \sim10^{14}$~M$_\sun$
\citep[e.g.,][]{iscs_paper,papovich_2010,brodwin2011,stanford2012,zeimann_2012,galametz12,muzzin2013,carla2013}.
The limiting factor for \spitzer ~is the relatively limited areas \citep[e.g., $\sim100$~deg$^2$,][]{ashby2013} 
for which it is feasible to obtain moderate-depth IRAC imaging.  The Wide-field Infrared Survey
Explorer mission \citep[WISE;][]{wise_paper} offers the means to find
massive clusters at $z>1$ over the full extragalactic sky using
methods pioneered with the IRAC camera aboard the {\it Spitzer Space
Telescope}.  We are conducting the next generation high-redshift
cluster search using WISE, the Massive and Distant Clusters of WISE
Survey (MaDCoWS).  We have already published the first $z \sim 1$
cluster confirmed from \madcows\ \citep{gettings2012}.  Here we report
on the initial results of follow-up observations to confirm an initial
sample of WISE-selected galaxy clusters selected to be at $z \sim 1$.
Brodwin et al. (in prep) will present Sunyaev-Zel'dovich (SZ)
measurements and cluster masses for some of the clusters in this
paper.


\section{WISE Selection of Cluster Candidates}
The key aspect of the WISE mission enabling detection of high-redshift galaxy clusters is the excellent photometric sensitivity at 3.4 and 4.6~$\mu$m (W1 and W2, respectively).  The WISE All-Sky Data Release achieves 5$\sigma$ sensitivity limits better than 0.07 and 0.1 mJy in unconfused regions in W1 and W2.\footnote{\url{http://wise2.ipac.caltech.edu/docs/release/allsky/expsup/sec6_3a.html}} 
These sensitivities correspond to areas near the ecliptic plane and represent the typical minimum depth for the All-Sky survey.  The effective exposure times of the observations vary significantly with ecliptic latitude due to the survey design, increasing from twelve 7.7s exposures at the ecliptic plane to more than a hundred near the ecliptic poles.  An overview of the methodology was given in \citet{gettings2012}, and full details of the \madcows\ galaxy cluster search method will be presented in a future paper; here we summarize. 

The search method is based on WISE All-Sky Release data, augmented by $i$-band photometry from the Sloan Digital Sky 
Survey (SDSS).  At 3.4~$\mu$m, the apparent magnitude of an $L^*$ galaxy is only weakly dependent upon redshift at $z 
\gtrsim1$ due to evolutionary and k-corrections which offset the impact of increasing luminosity distance.
  Consequently, a magnitude-limited galaxy sample selected at 3.4~$\mu$m has a nearly uniform luminosity limit at this epoch.   This is even more true for the 4.6$\mu$m band (Eisenhardt et al. 2008), but the All-Sky WISE release is not deep enough to select at W2.  
  In contrast, this redshift dependence increases in strength at progressively shorter wavelengths, enabling color selection of distant galaxies.

  The \madcows{} search exploits these properties using a set of magnitude and color cuts in WISE and optical bands to isolate a population of $z\gtrsim 1$ galaxies. 
  After cleaning the WISE catalog of flagged sources and applying a 3.4~$\mu$m cut to ensure spatial uniformity in depth, we match the WISE catalog to the SDSS DR8 photometric catalog \citep{dr8_paper}.
  In the current version of the \madcows{} search, which covers 15 of the 19 clusters in this paper, we then reject sources with $\rm{W}1-\rm{W}2 < 0.2$ and $i<21$, effectively removing the foreground galaxy population. 
  The other four clusters presented here are from a preliminary search which employed the same technique but lacked the current optimization of these color and magnitude cuts.
  Once foreground sources are removed, we input the remaining sources to a wavelet search to identify overdensities on a scale of ${\sim}3^{\prime}$, which corresponds to a physical size of ${\sim}1.4$~Mpc at $z\sim1$. 
  From the resulting wavelet-smoothed density maps we select the most-significant overdensities as MaDCoWS candidate clusters.

The remaining sources are smoothed with a Gaussian-difference wavelet kernel tuned to enhance structures on scales of $\sim 3'$, corresponding to a physical size of $\sim1.4$~Mpc at $z\sim1$.  From these maps we identified the most statistically significant overdensities for further investigation.  The centroids of the detections are used below to define the cluster center.  A more detailed description of the creation of the MaDCoWS cluster catalog will be presented in a forthcoming publication (Gettings et al., in preparation)


\section{Follow-Up Observations}
\label{followup}
As described below, we have been carrying out follow-up observations to confirm WISE-selected candidate clusters.  Images of the spectroscopically confirmed candidates are shown in Figure 1, and listed in Table~\ref{observations_table}.  Optical imaging was obtained with the Gran Telescopio Canarias (GTC) and Gemini-North, Infrared Array Camera (IRAC; \citealt{irac}) imaging with the $Spitzer~Space~Telescope$, and optical spectroscopy with Keck and Gemini-North, as described below.   In general, cluster candidates were chosen for optical spectroscopic confirmation based primarily on three factors:  the appearance of a red sequence in the color-magnitude diagram, their visual appearance in the optical imaging, and the galaxy overdensity as measured in the IRAC images. 

\subsection{GTC Imaging}
\label{followup:gtc}

We obtained $i$-band imaging for five of the candidate clusters in Table 1 using the optical imaging instrument OSIRIS on the GTC in queue mode in August 2012.  
A total of 18 minutes of exposure time was obtained on each of these clusters, using six dithered 180 second integrations. 
The data were processed using standard reduction procedures in IRAF and astrometrically calibrated to the USNO-B1.0 catalog \citep{usnob_paper} using SCAMP and SWarp \citep{scamp, swarp}.   The seeing in these optical images is approximately $0.7 - 0.8$ arcsec.  
Photometric calibration of final, stacked images was performed using DR10 \citep{sdss3} photometry of SDSS stars in each field. 

\subsection{Gemini-North Imaging}

As part of a queue program at Gemini-North (Brodwin, PI) which includes spectroscopy, we obtained GMOS $r$ and $z$-band imaging in the 2013A and 2013B semesters on 12 of the candidates listed in Table~1.  A series of dithered exposures was obtained in each band, $5 \times 180$~s in the $r$-band, and $12 \times 80$~s in the $z$-band.  
The data were processed using standard reduction procedures in IRAF and astrometrically calibrated to the USNO-B1.0 catalog \citep{usnob_paper} using SCAMP and SWarp.  The seeing in the GMOS images is typically $0.6$ arcsec.   Photometric calibration was obtained using DR10 photometry of SDSS stars that appear in each field.  


\subsection{IRAC Imaging}

As part of a larger $Spitzer$ survey of MaDCoWS candidates (Gonzalez, PI), we obtained IRAC $3.6$ and $4.5$ $\mu$m imaging in Cycle 9 for 200 MaDCoWS cluster candidates including all of the clusters in Table 1; here we make use of the IRAC data for only 7 of the candidates listed in Table~1 in the cases where our optical imaging is either incomplete or nonexistent.   A series of six dithered 30~s exposures was obtained with IRAC in each channel for a total exposure time of $180$~s.  The basic calibrated data were reduced and mosaiced using MOPEX \citep{mopex} and resampled to a pixel scale of $0.6$ arcsec.  The PSF in the IRAC images has a FWHM $\sim 1.7$ arcsec.   A more detailed presentation of the MaDCoWS IRAC program will be made in a future publication.

\subsection{Keck Spectroscopy}
\label{followup:keck}

The optical/infrared catalogs described below in Section~\ref{cmdtext} were used to design slit-masks for use with DEIMOS at the W. M. Keck Observatory to obtain redshifts of potential cluster galaxies.  For clusters where Gemini $r-z$ photometry was available, we used the red sequence as primary targets,  whereas for clusters with GTC imaging, we used $i-W1$ colors to select targets, and for the clusters with only $Spitzer$ imaging, we selected on the basis of [3.6]$-$[4.5] color. 
The general approach to designing the slit masks was to rely on the red sequence as the source of the highest priority objects, weighting by cluster-centric radius, and then filling in the masks with other galaxies at larger radii selected by WISE W1$-$W2 color.  
The slitlets in the masks had widths of $1\farcs1 $ and minimum lengths of 5 arcsec.   Spectra were obtained using the 600 line grating and the GG495 filter on three observing runs in the 2012B, 2013A, and 2013B semesters, when conditions were generally clear with seeing of $0\farcs6 - 1\farcs0$.  Table~1 lists the observation dates for each cluster.    
For most masks we obtained three 1200~s exposures, with longer exposure times being devoted to the higher redshift candidates. The DEIMOS spectra were reduced using the DEEP2 pipeline.   
The relative spectral response of the extracted spectra was calibrated via observations of Feige 34 and BD 28+4211.

\subsection{Gemini-N Spectroscopy}

Using the optical catalogs created from GMOS preimaging, we designed slit masks for GMOS to obtain redshifts of potential cluster galaxies, following a similar procedure to design masks. Spectra were obtained using $1\farcs0 $ slit widths, the R400 grating, and the RG810 filter.  Three sets of nod and shuffle sequences were obtained at each of two central wavelength settings (8100 and 8200 Angstroms).  Each nod and shuffle sequence used $\pm 0.75$ arcsec nods, with 9 cycles of 60~s exposures, so the total on-source exposure time was 6480~s.  The seeing was in the $0\farcs6$ to $1\farcs0$ range.  Table~1 lists the observation dates for each cluster.  The GMOS spectra were reduced using standard procedures in IRAF.      

\section{Results}
\label{results}

\subsection{Color-magnitude Diagrams}\label{cmdtext}

For all of the imaging described above, SExtractor \citep{sextractor} catalogs were generated using the longer wavelength image for detection.  Photometry was obtained from the shorter wavelength image using SExtractor in dual band mode.  Magnitudes were measured using circular apertures with a diameter of 2.5 arcsec, except for the clusters where we use IRAC images for which the aperture size is 4 arcsec diameter, and were aperture-corrected.  All magnitudes are on the AB system.  

Color magnitude diagrams (CMDs) were constructed for all the confirmed clusters in Table~1, and are presented in Figure 2.   For MOO~1625+2629 and MOO~1210+3154 we do not have any optical imaging, so the CMDs use [3.6] $-$ [4.5] colors, which have a more limited power to discriminate the red sequence of a $z \sim 1$ cluster from field galaxies.  For reference, the luminosity and color of an $L^*$ galaxy appropriate to each cluster's redshift are shown in the CMDs.  The color was calculated from a passive evolution model using EzGal \citep{ezgal} with the following parameters:  the 2007 version of a \citet{bc03} simple stellar population (SSP) model, $z_f = 3$, Solar metallicity, and a Chabrier (2003) IMF.  The luminosity is calculated using a normalization of the Bruzual-Charlot model to the Coma cluster luminosity function \citep{coma}.  The spectroscopic member galaxies are marked by red circles, and objects within the red sequence are marked by yellow squares.  The red sequence is defined as magnitudes within one magnitude of L$^*$, and 0.3 magnitudes of the model color.

\subsection{Redshifts}
\label{results:redshifts}

Redshifts of objects in each cluster were determined by visual inspection of the reduced spectra and identification of prominent spectral features including the 4000 Angstrom break, the Ca H+K absorption lines, and the [\ion{O}{2}] $\lambda 3727$ emission line.  Sample spectra are shown in Figure~\ref{spec} for one of the lower-redshift clusters and for one of the higher redshift clusters.  Redshift information for all the cluster members is given in Table~\ref{redshifts}.  The spectroscopic member galaxies are marked in the color-magnitude diagrams as well as in the optical/infrared images shown in Figures 1 and 2, respectively.  Clusters for which at least 5 objects have redshifts within $\pm 2000 (1 + z)$ km~s$^{-1}$ and within a radius of 2 Mpc are considered to be confirmed.  

The overall results of the spectroscopy are summarized in Table~\ref{clusters}, which lists the confirmed clusters with their mean redshifts, and in Figure~\ref{zhist}.  The redshift histogram shows that the MaDCoWS selection technique effectively identifies $z \sim 1$ galaxy clusters.  Four of the new clusters, represented in Figure~\ref{zhist} by the open areas of the histogram, were selected with a preliminary version of the search criteria but are not in the final selection of cluster candidates.  They have a redshift distribution broadly similar to the final selection.  

The spectroscopic results can be used to make a preliminary assessment of the purity of the sample, with the caveat that the selection method has evolved over the course of the spectroscopic campaigns, as both the spectroscopic results and IRAC imaging were used to refine the cluster search.  To assess the purity of the MaDCoWS sample, we consider the confirmation rate of cluster candidates that are among the 200 highest significance detections in the current iteration of the WISE search.  From this list, we have spectroscopically confirmed 14 clusters.  In addition to these, we obtained optical spectroscopy of 6 other candidates in the current MaDCoWS selection.  For one candidate the redshifts show the presence of multiple groups along the line of sight and so we deem this candidate to be a failure.  For the remaining five candidates, we find a spread of redshifts with no definitive evidence of groups or clusters.  Because only one mask was obtained on each of these five candidates, we cannot yet confidently state whether they are real or spurious.  Considering the limiting cases where all or none of these five candidates are spurious, we find that the purity lies in the range 70-95\%.

\section{Discussion}
\label{discussion}

We have presented imaging and spectroscopy that confirms 19 new galaxy clusters in the $0.7 < z < 1.3$ range, which were selected using the All-Sky WISE data release.  This set of new clusters, which includes nine at $z > 1$, robustly demonstrates the viability of conducting an all-sky search for such objects using WISE.   The distribution of the cluster redshifts indicates that our final MaDCoWS selection technique identifies galaxy clusters primarily in the $0.8 < z < 1.3$ range.  To date, five of the clusters in Table 3 have been observed by CARMA and have measured SZ decrements, consistent with M$_{200}$ in the range $2 - 6 \times 10^{14}$~M$_\odot$ range (Brodwin et al., in prep.).   As a result, the clusters identified by MaDCoWS are expected to form a valuable sample for obtaining cosmological constraints, and for better understanding of the formation of massive clusters and their constituent galaxy populations at $z > 1$.   

The color-magnitude diagrams, optical images, and spectroscopy suggest that the MaDCoWS clusters span a variety of cluster types.  Some clusters such as MOO~1514$+$1316 appear to be relaxed, with a centrally concentrated core of galaxies, a well-defined red sequence, and few emission-line objects.  But other clusters span the range of possible combinations of these characteristics, some having few emission-line objects but little if any spatial concentration in a core region, and others having a relatively high fraction of emission-line galaxies and a well-formed cluster core.  The optical spectroscopy was being carried out while we were fine-tuning the WISE selection procedure, and this could contribute to the variety of cluster properties in the confirmed clusters.  As the number of confirmed clusters and members in each cluster increases, it will become possible to draw more robust conclusions about the galaxy populations in MaDCoWS clusters.

The prospects for extending the redshift reach, the upper mass range, and the effective area of MaDCoWS are excellent.  Some of the newly confirmed clusters were chosen for follow-up without the benefit of seeing the IRAC images, which were obtained later than the observing runs at the ground-based telescopes.  Now that we can fully exploit our IRAC imaging of the 200 richest WISE cluster candidates to select targets for future follow-up at e.g. CARMA and Keck, we expect that the mean cluster mass of the MaDCoWS cluster sample will rise.  Additionally, the clusters presented here were discovered using the original All-Sky WISE data.  The AllWISE data release (Cutri et al. 2013), which contains twice the exposure time in the W1 and W2 bands, now provides deeper photometry in the W1 and W2 bandpasses, enabling a search extending to higher redshifts which we are now undertaking.  In the future, the prospect of combining the one year of AllWISE survey data with the three years of additional survey data expected from the recently reactivated NEOWISE survey (Mainzer et al.  in prep) should allow us to select at W2 and identify robust cluster candidates to  $z > 1.5$.  Finally, the area so far covered by MadCoWS has been restricted to that of the SDSS, as their optical imaging efficiently suppresses low-redshift interlopers that would otherwise be selected by the WISE W1$-$W2 color.  As more area is covered by optical imaging surveys of similar quality, such as the Dark Energy Survey, we will expand the area of our WISE search for the most massive galaxy clusters at $z \sim 1$ and above.

\acknowledgements
This publication makes use of data products from the Wide-field Infrared Survey Explorer, which is a joint project of the University of California, Los Angeles and the Jet Propulsion Laboratory/California Institute of Technology, funded by the National Aeronautics and Space Administration (NASA).
Funding for SDSS-III has been provided by the Alfred P. Sloan Foundation, the Participating Institutions, the National Science Foundation, and the U.S. Department of Energy Office of Science. The SDSS-III web site is http://www.sdss3.org/.
SDSS-III is managed by the Astrophysical Research Consortium for the Participating Institutions of the SDSS-III Collaboration, a list of which can be found at https://www.sdss3.org/collaboration/institutions.php.
We acknowledge using EzGal, available at www.baryons.org/ezgal/index.php, to calculate the colors displayed in the color-magnitude diagrams.
S.A.S, M.B., D. P. G. and A. H. G. acknowledge support for this research from the NASA Astrophysics Data Analysis Program (ADAP) through grant NNX12AE15G. 
Some of the data presented herein were obtained at the W.M. Keck Observatory, which is operated as a scientific partnership among the California Institute of Technology, the University of California and the National Aeronautics and Space Administration. The Observatory was made possible by the generous financial support of the W.M. Keck Foundation. 
A.H.G and D. P. G. were Visiting Astronomers at Gemini Observatory, National Optical Astronomy Observatory, which is operated by the Association of Universities for Research in Astronomy (AURA) under cooperative agreement with the National Science Foundation. 
This work is based in part on observations made with the $Spitzer~Space~Telescope$, which is operated by the Jet Propulsion Laboratory, California Institute of Technology, under a contract with NASA.
Based on observations made with the Gran Telescopio Canarias (GTC), installed in the Spanish Observatorio del Roque de los Muchachos of the Instituto de Astrof'sica de Canarias, on the island of La Palma.
We thank the anonymous referee for comments which improved the final manuscript. 

\bibliographystyle{apj}

\begin{deluxetable}{lllll}
\tablewidth{0pt}
\tabletypesize{\scriptsize}
\tablecaption{Follow-up Observations}
\tablehead{ \colhead{Cluster ID} & \colhead{$\alpha$ (J2000)}& \colhead{$\delta$ (J2000)} & \colhead{Imaging\tablenotemark{2}} & \colhead{Spectroscopy--Instrument: UT Dates} }
\startdata
MOO\tablenotemark{1}J0012$+$1602 & 00:12:13.0  & $+$16:02:16 & GMOS & DEIMOS: 2013-09-08, 2013-09-09 \\
MOO~J0024$+$3303 & 00:24:44.9 & $+$33:03:10 & GMOS & GMOS: 2013-11-25, DEIMOS: 2013-09-09 \\
MOO~J0125$+$1344 & 01:25:22.4 & $+$13:44:35 & GMOS & GMOS: 2013-11-26, DEIMOS: 2013-09-08 \\
MOO~J0130$+$0922 & 01:30:36.4 & $+$09:22:36 & GTC, IRAC & DEIMOS:  2012-10-18, 2012-10-19 \\
MOO~J0133$-$1057 & 01:33:55.6 & $-$10:57:44 & GTC, IRAC & DEIMOS: 2012-10-18, 2012-10-19 \\
MOO~J0212$-$1813 & 02:12:04.1 & $-$18:14:13 & GTC, IRAC & DEIMOS: 2012-10-18  \\
MOO~J0224$-$0620 & 02:24:51.3 & $-$06:20:17 & GMOS & GMOS: 2013-12-04  \\
MOO~J0245$+$2018 & 02:45:07.9 & $+$20:18:15 & GMOS & GMOS: 2012-10-16+17, DEIMOS: 2012-10-18+19 \\
MOO~J0319$-$0025 & 03:19:25.6  & $-$00:25:20 & GMOS & DEIMOS: 2013-09-07 \\
MOO~J1155$+$3901 & 11:55:45.4  & $+$39:01:06 & GMOS & GMOS: 2013-05-28 \\
MOO~J1210$+$3154 & 12:10:54.3 & $+$31:54:40 & IRAC  & DEIMOS: 2013-05-09 \\
MOO~J1319$+$5519 & 13:19:38.3 & $+$55:19:07 & GMOS & GMOS: 2013-06-30, 2013-07-01 \\
MOO~J1335$+$3004 & 13:35:41.7 & $+$30:04:13  & GMOS & GMOS: 2013-04-19 \\
MOO~J1514$+$1346 & 15:14:43.8 & $+$13:46:32 & GMOS & GMOS: 2013-04-19, DEIMOS: 2013-07-13 \\
MOO~J1625$+$2629 & 16:25:03.4 & $+$26:29:54 & IRAC & DEIMOS: 2013-09-09 \\
MOO~J2205$-$0917 & 22:05:37.5 & $-$09:17:28 & GMOS & GMOS: 2013-10-27 \\
MOO~J2320$-$0620 & 23:20:16.8 &$-$06:20:32 & GTC, IRAC & DEIMOS: 2012-10-18 \\
MOO~J2348$+$0846 & 23:48:19.9 & $+$08:46:01 & GTC, IRAC & DEIMOS: 2012-10-13, 2012-10-19 \\
MOO~J2355$+$1030 & 23:55:01.0 & $+$10:30:43 & GMOS & DEIMOS: 2013-09-09 \\
\enddata
\tablenotetext{1}{MOO stands for MaDCoWS Overdense Object}
\tablenotetext{2}{Source of imaging used to create color-magnitude diagrams in Figure 2.}

\label{observations_table}
\end{deluxetable}

\clearpage

\makeatletter 
\renewcommand{\thefigure}{\@arabic\c@figure a}
\makeatother

\begin{figure}
\includegraphics[width=0.49\linewidth]{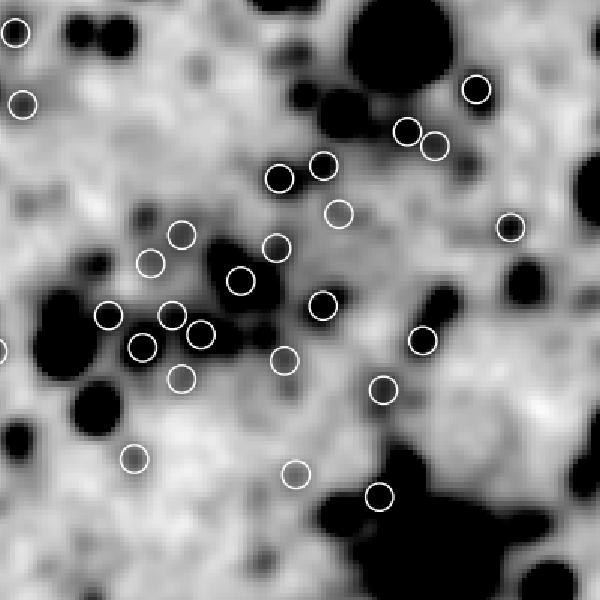}
\hfill
\includegraphics[width=0.49\linewidth]{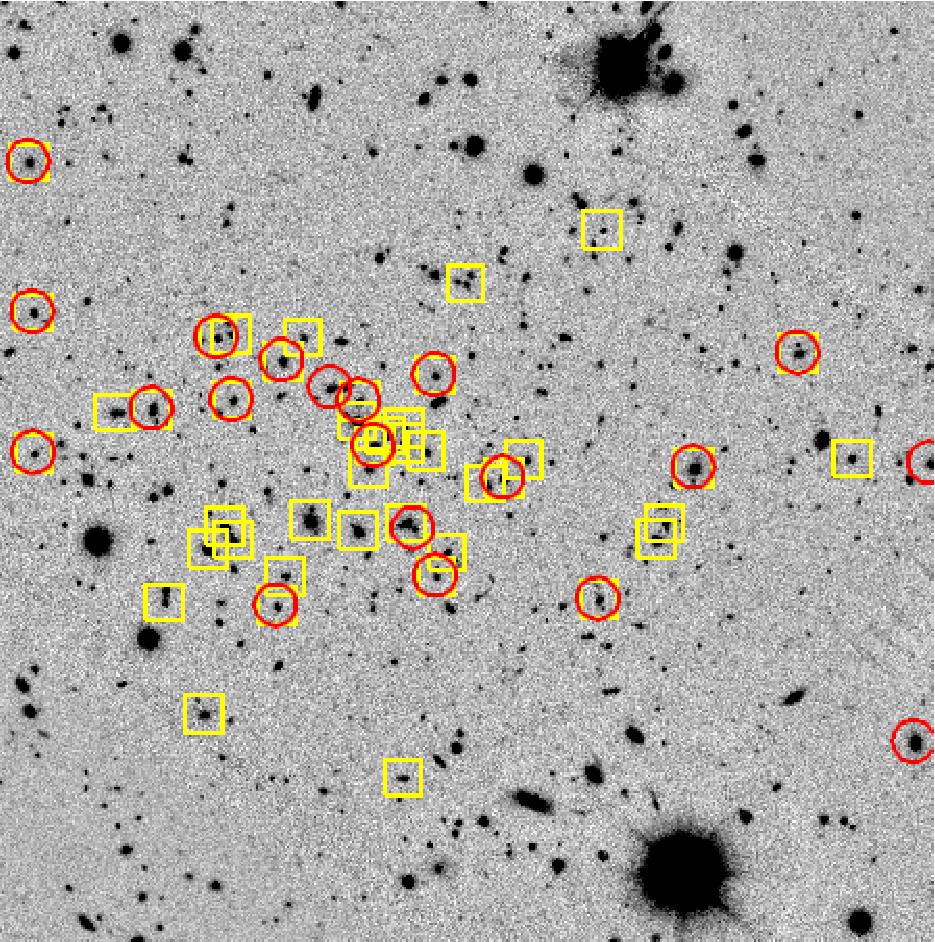}
 \caption{MOO~0012$+$1602, $z=0.94$, \emph{left:}  $3^\prime\times 3^\prime$ WISE W1 image. The white circles represent sources identified as probable high-redshift galaxies based upon color and magnitude cuts (see text).  The white circles have a diameter of 8.1\arcsec, similar to the full width at half maximum of the WISE W1 and W2 Atlas images.  North is up and East left for all images.
   \emph{Right:} $3^\prime\times 3^\prime$ $z$-band image of the same area as in the left panel.  Red circles mark the spectroscopically confirmed members (some of which lie outside the FOV of the image) and the yellow boxes mark the objects in the red sequence (see text). }
 \label{fig:m0012p1602img}
\end{figure}

\setcounter{figure}{0}
\makeatletter 
\renewcommand{\thefigure}{\@arabic\c@figure b}
\makeatother

\begin{figure}
\includegraphics[width=0.49\linewidth]{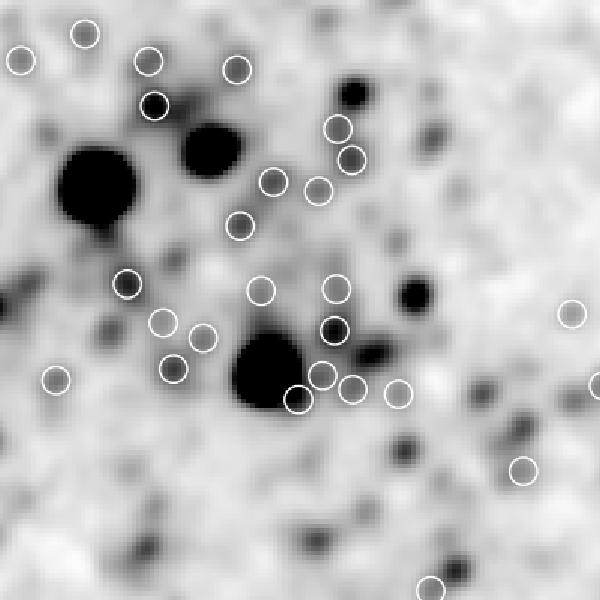}
\hfill
\includegraphics[width=0.49\linewidth]{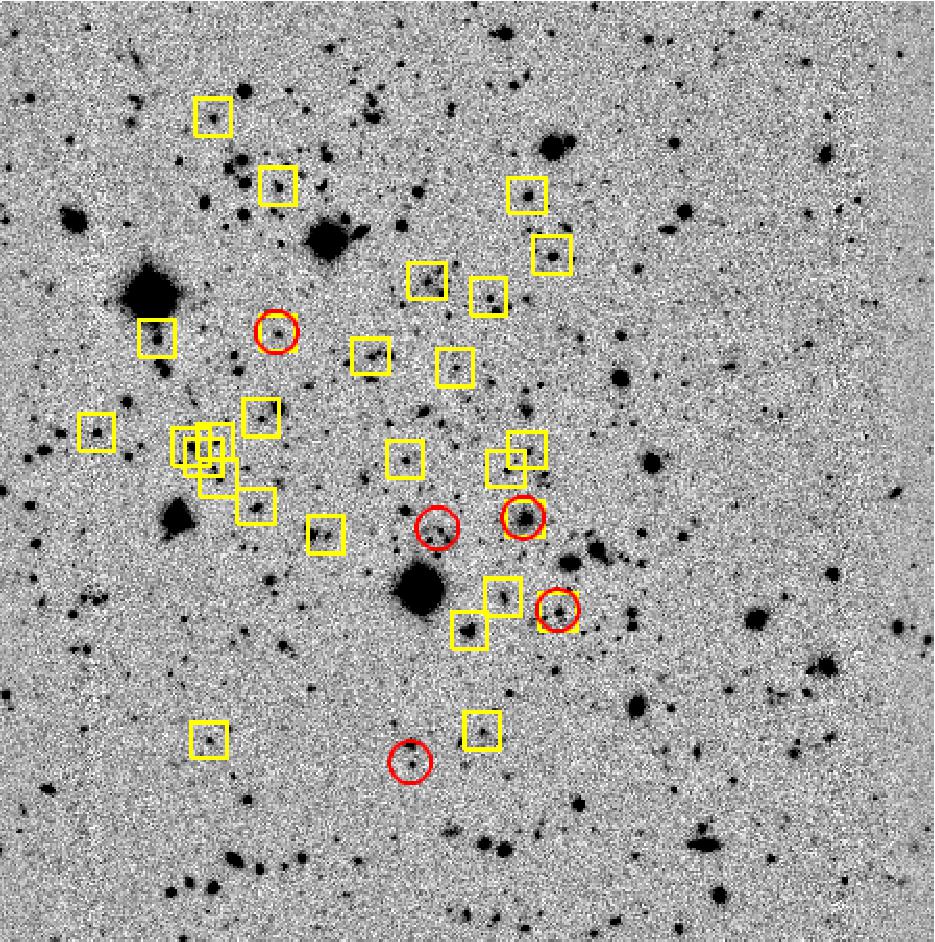}
 \caption{MOO~0024$+$3303, $z = 1.11$, as above; W1 image on the left, $z$-band image on the right.
 }
 \label{fig:m0024p3303img}
\end{figure}

\setcounter{figure}{0}
\makeatletter 
\renewcommand{\thefigure}{\@arabic\c@figure c}
\makeatother

\begin{figure}
\includegraphics[width=0.49\linewidth]{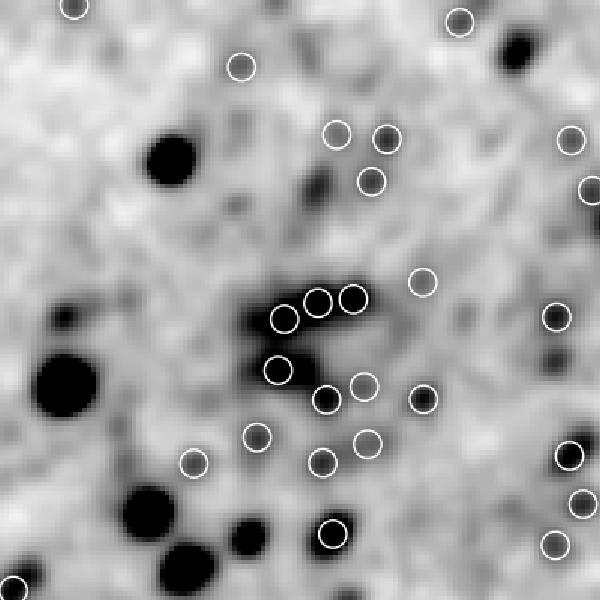}
\hfill
\includegraphics[width=0.49\linewidth]{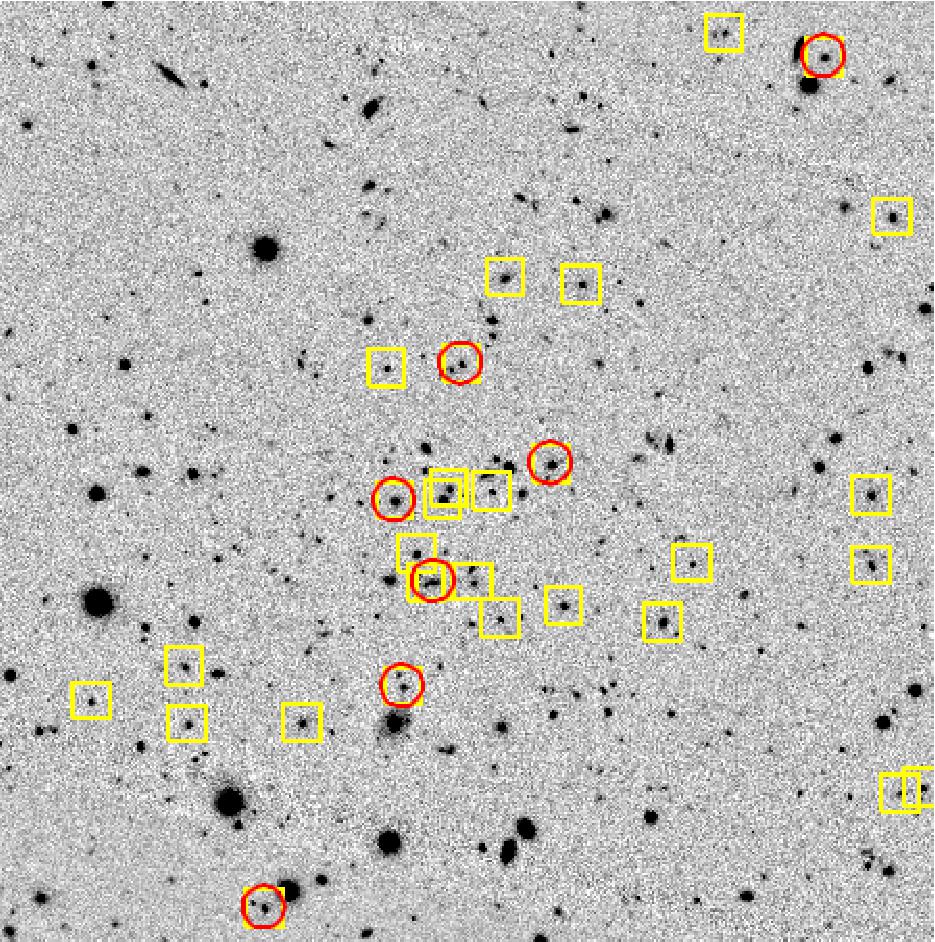}
 \caption{MOO~0125$+$1344, $z = 1.12$, as above; W1 image on the left, $z$-band image on the right.
 }
 \label{fig:m0125p1344img}
\end{figure}

\setcounter{figure}{0}
\makeatletter 
\renewcommand{\thefigure}{\@arabic\c@figure d}
\makeatother

\begin{figure}
\includegraphics[width=0.49\linewidth]{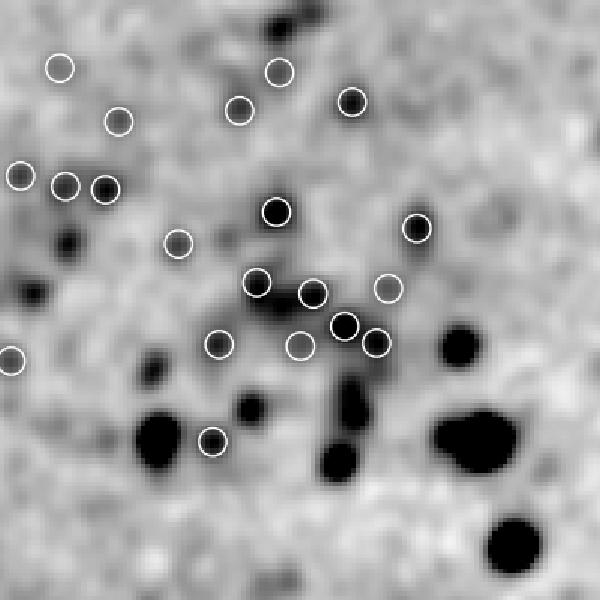}
\hfill
\includegraphics[width=0.49\linewidth]{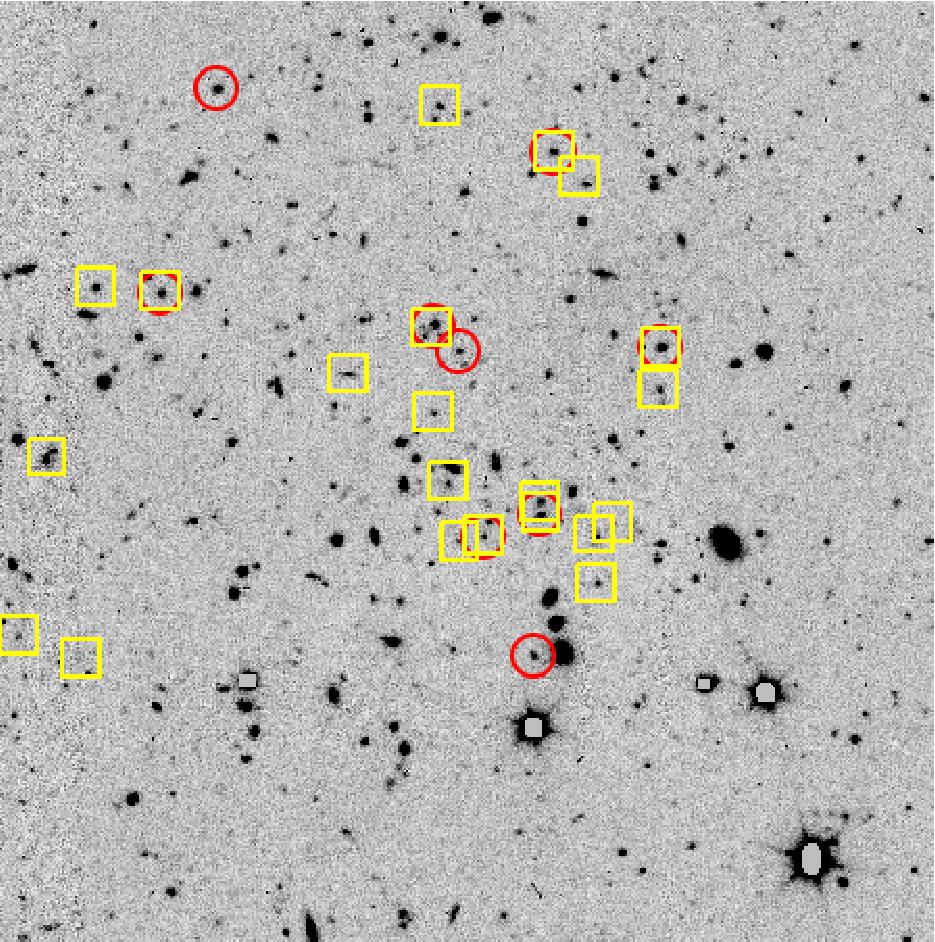}
 \caption{MOO~0130$+$0922, $z = 1.15$, as above; W1 image on the left, $i$-band image on the right.
 }
 \label{fig:m0130p0922img}
\end{figure}

\setcounter{figure}{0}
\makeatletter 
\renewcommand{\thefigure}{\@arabic\c@figure e}
\makeatother

\begin{figure*}
\includegraphics[width=0.49\linewidth]{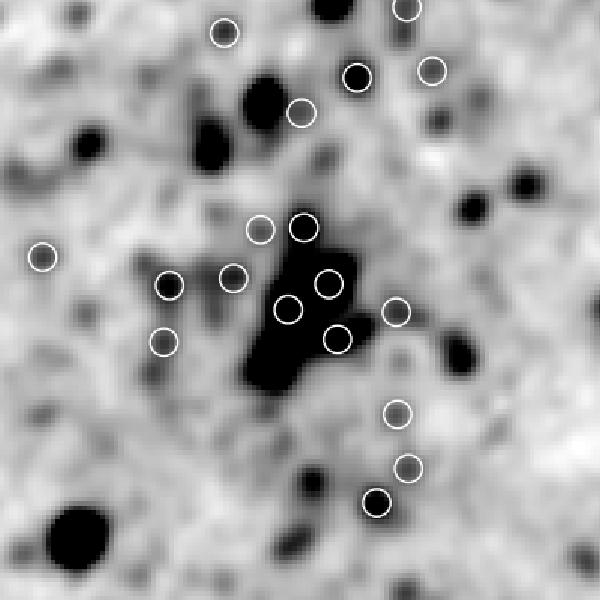}
\hfill
\includegraphics[width=0.49\linewidth]{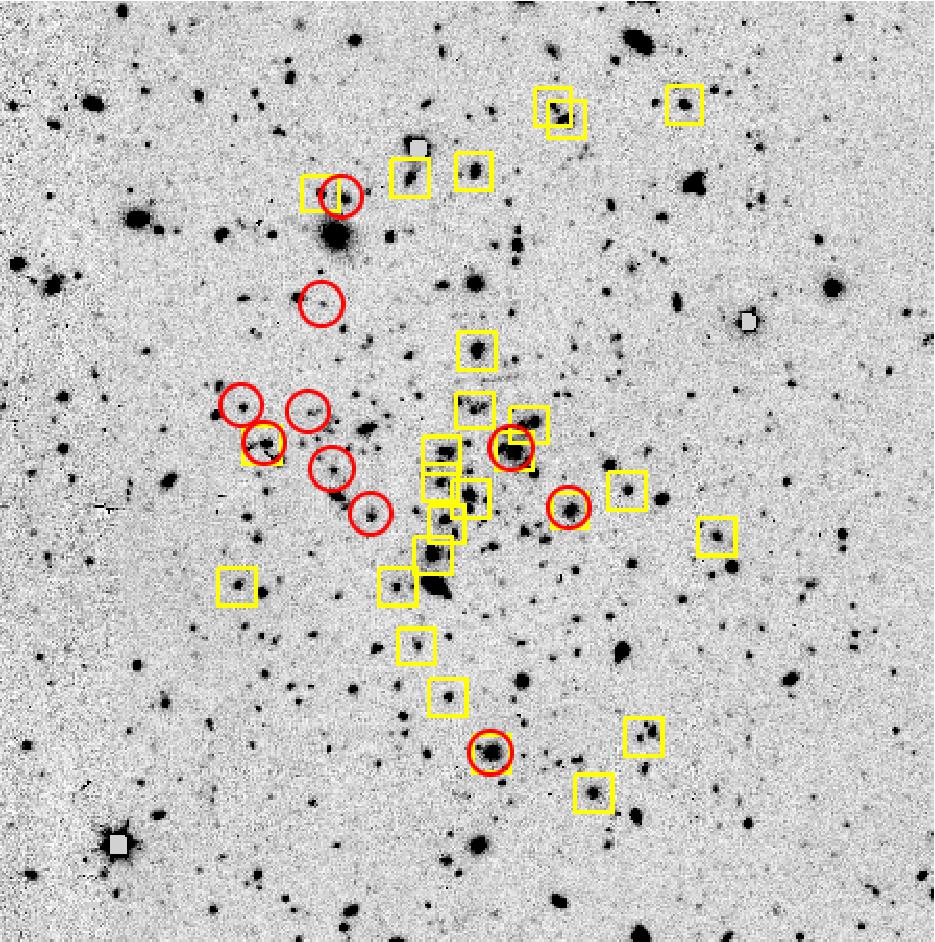}
 \caption{MOO~0133$-$1057, $z = 0.96$, as above; W1 image on the left, $i$-band image on the right.
 }
 \label{fig:m0133m1057img}
\end{figure*}

\setcounter{figure}{0}
\makeatletter 
\renewcommand{\thefigure}{\@arabic\c@figure f}
\makeatother

\begin{figure*}
\includegraphics[width=0.49\linewidth]{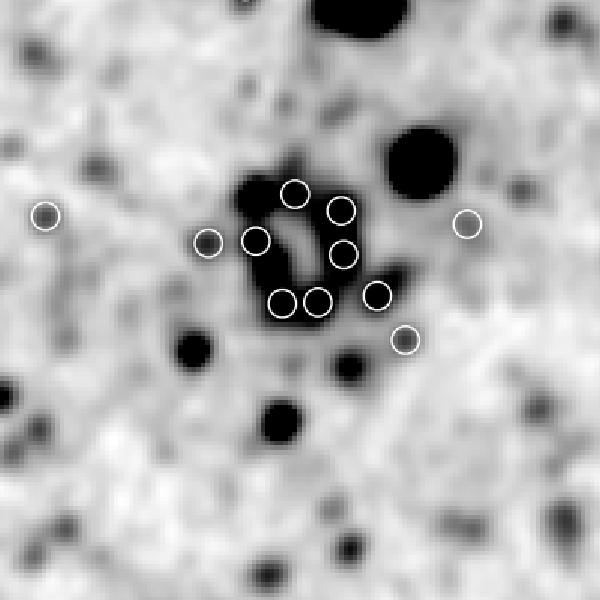}
\hfill
 \includegraphics[width=0.49\linewidth]{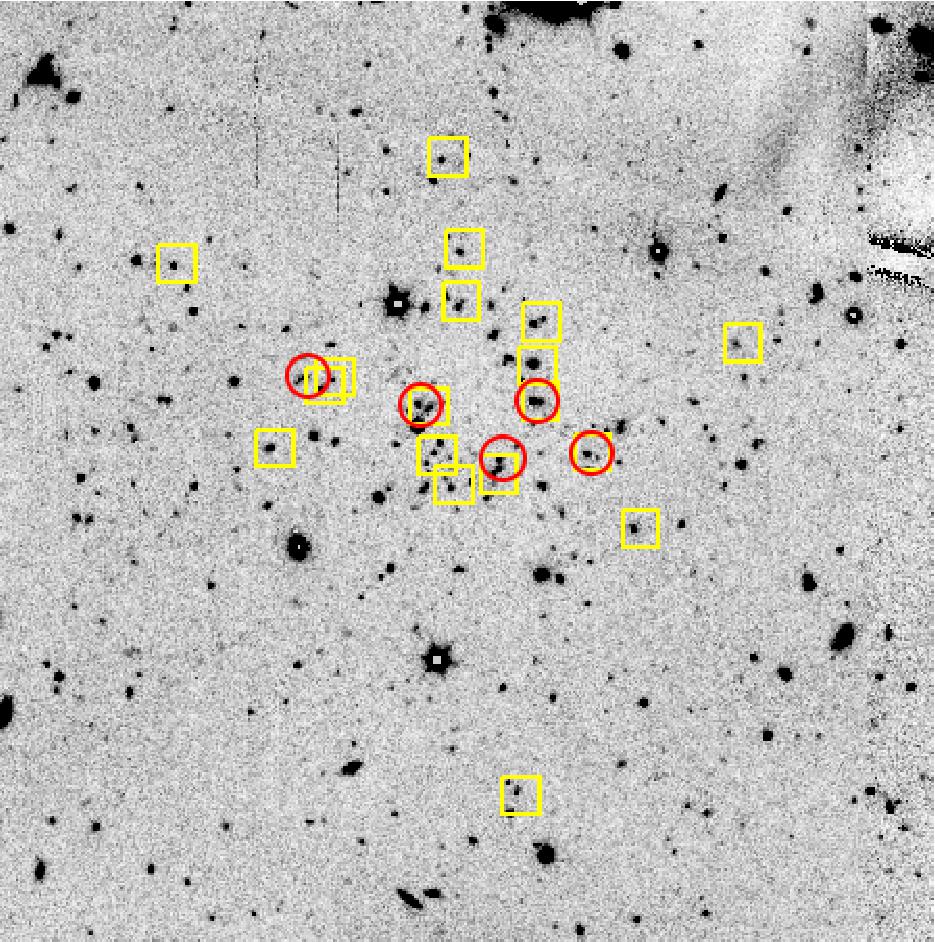}
 \caption{MOO~0212$-$1813, $z = 1.09$, as above; W1 image on the left, $i$-band image on the right.
 }
 \label{fig:m0212m1813img}
\end{figure*}

\setcounter{figure}{0}
\makeatletter 
\renewcommand{\thefigure}{\@arabic\c@figure g}
\makeatother

\begin{figure*}
\includegraphics[width=0.49\linewidth]{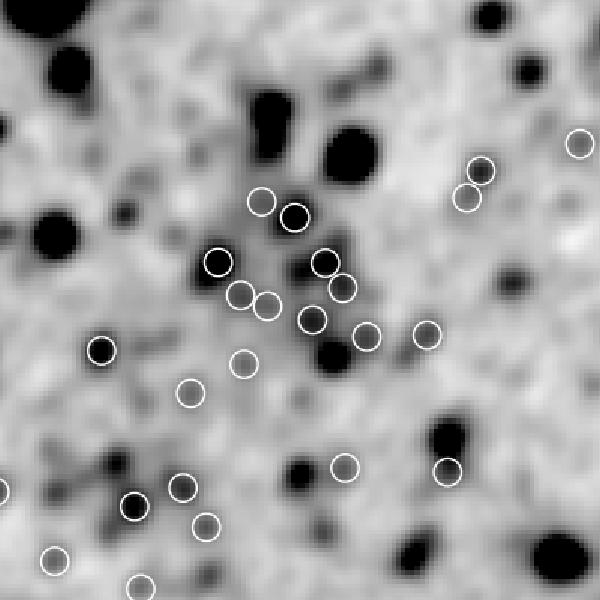}
\hfill
 \includegraphics[width=0.49\linewidth]{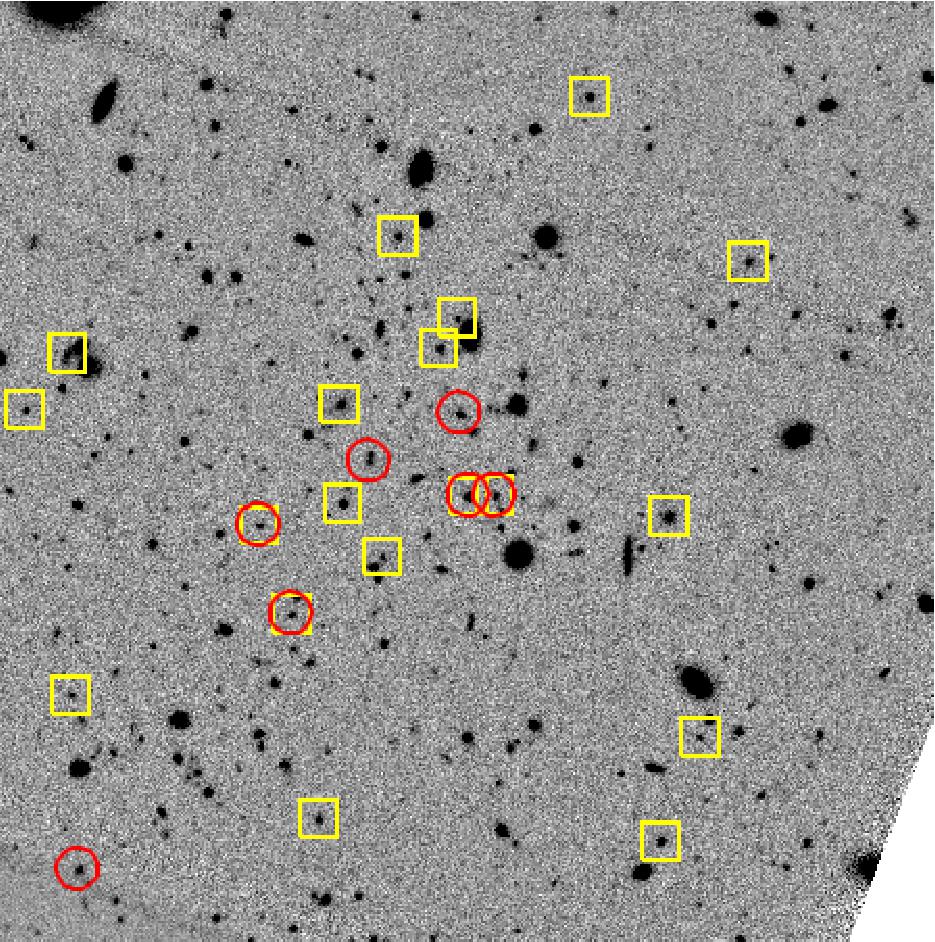}
 \caption{MOO~0224$-$0620, $z = 0.81$, as above; W1 image on the left, $z$-band image on the right.  
 }
 \label{fig:m0224m0620img}
\end{figure*}

\setcounter{figure}{0}
\makeatletter 
\renewcommand{\thefigure}{\@arabic\c@figure h}
\makeatother

\begin{figure*}
\includegraphics[width=0.49\linewidth]{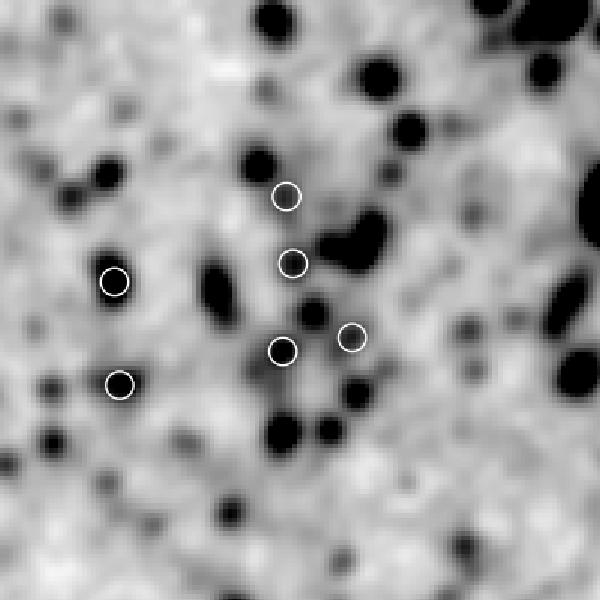}
\hfill
 \includegraphics[width=0.49\linewidth]{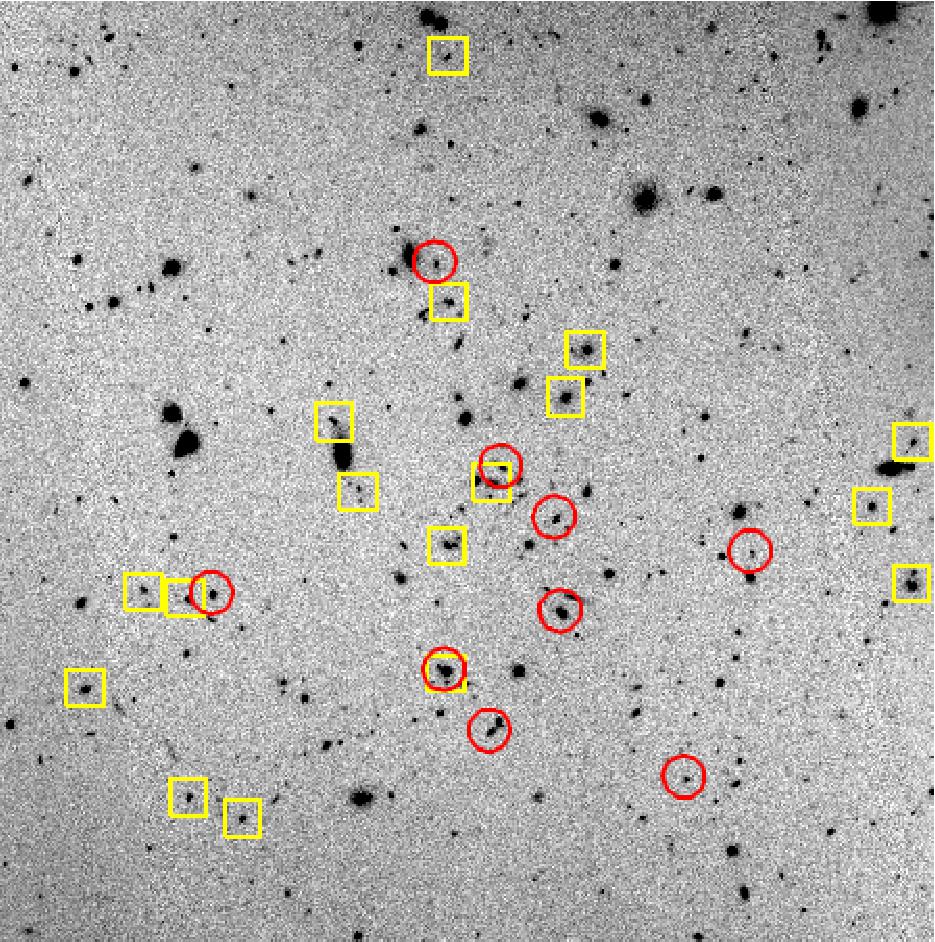}
 \caption{MOO~0245$+$2018, $z = 0.76$, as above; W1 image on the left, $i$-band image on the right.
 }
 \label{fig:m0245p2018img}
\end{figure*}

\setcounter{figure}{0}
\makeatletter 
\renewcommand{\thefigure}{\@arabic\c@figure i}
\makeatother

\begin{figure*}
\includegraphics[width=0.49\linewidth]{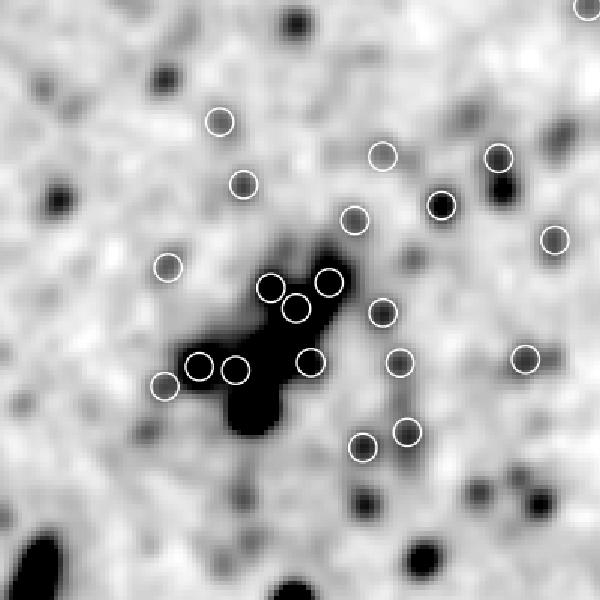}
\hfill
 \includegraphics[width=0.49\linewidth]{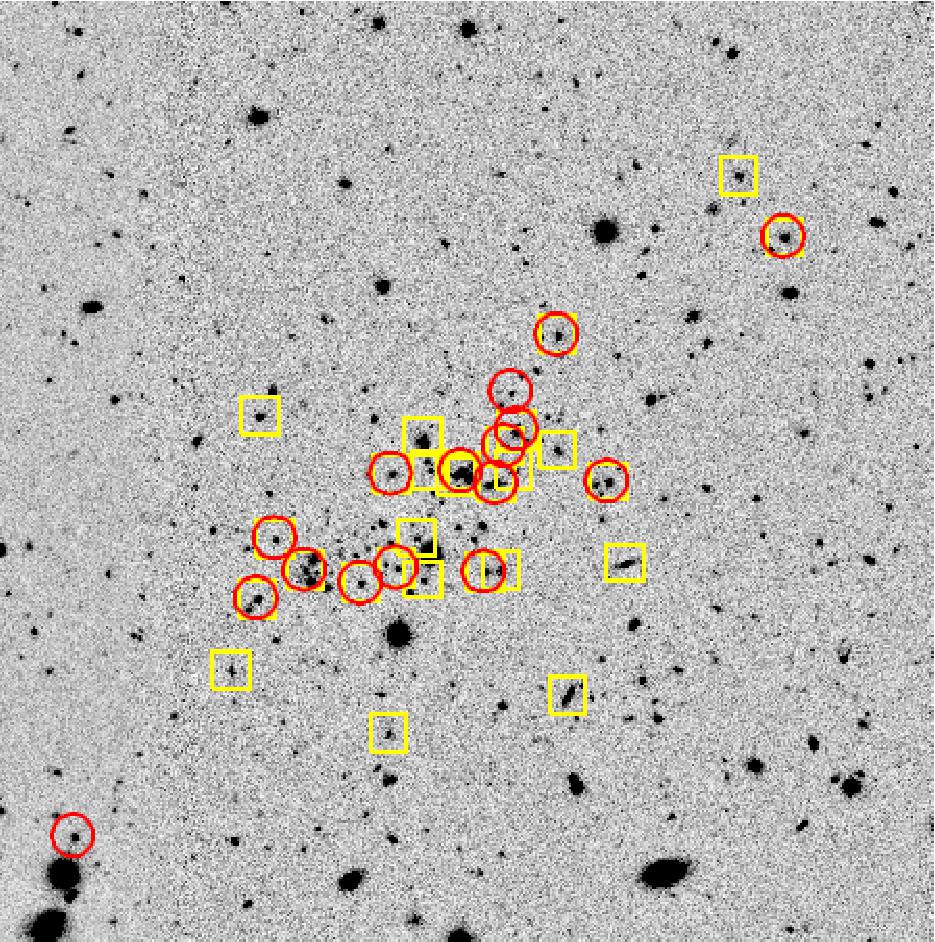}
 \caption{MOO~0319$-$0025, $z = 1.19$, as above; W1 image on the left, $z$-band image on the right.
 }
 \label{fig:m0319m0025img}
\end{figure*}

\setcounter{figure}{0}
\makeatletter 
\renewcommand{\thefigure}{\@arabic\c@figure j}
\makeatother

\begin{figure*}
\includegraphics[width=0.49\linewidth]{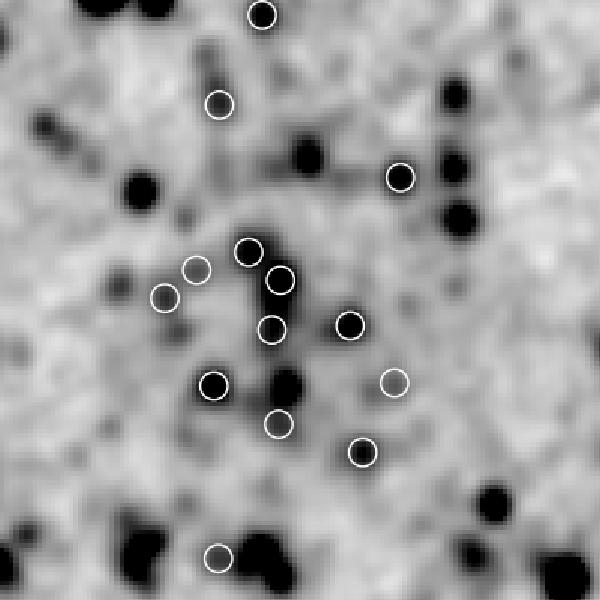}
\hfill
 \includegraphics[width=0.49\linewidth]{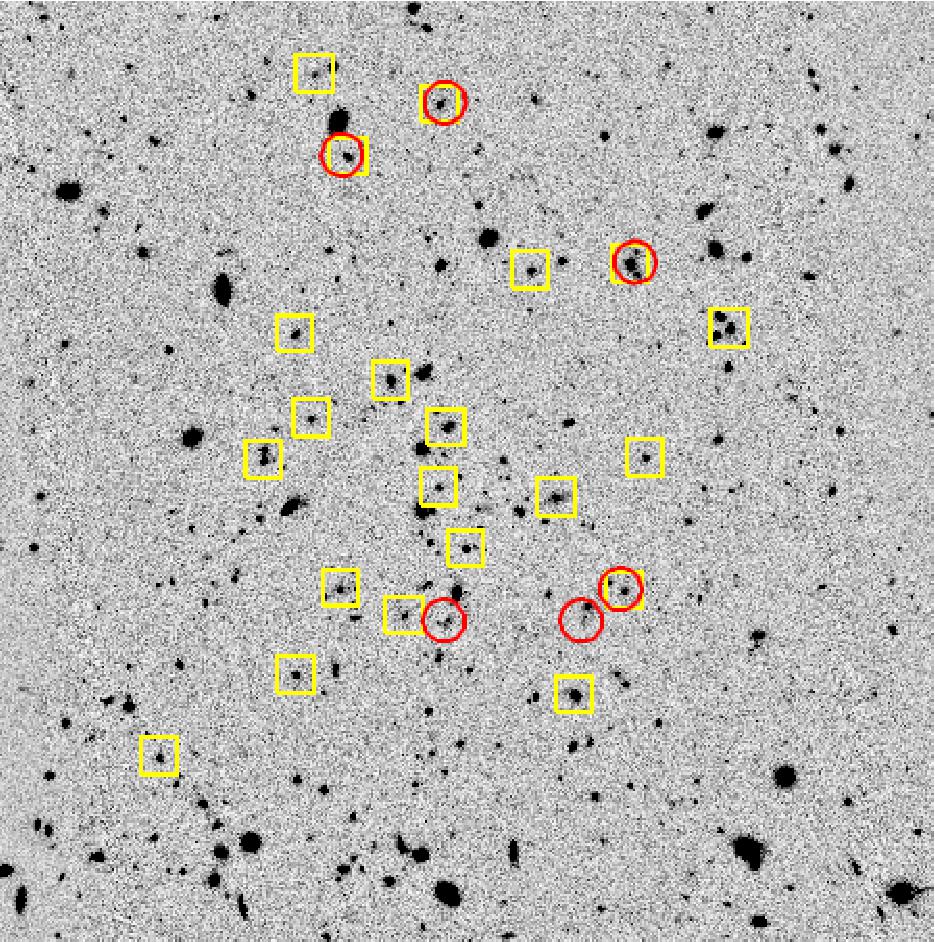}
 \caption{MOO~1155$+$3901, $z = 1.01$, as above; W1 image on the left, $z$-band image on the right.
 }
 \label{fig:m11155p3901img}
\end{figure*}

\setcounter{figure}{0}
\makeatletter 
\renewcommand{\thefigure}{\@arabic\c@figure k}
\makeatother

\begin{figure*}
\includegraphics[width=0.49\linewidth]{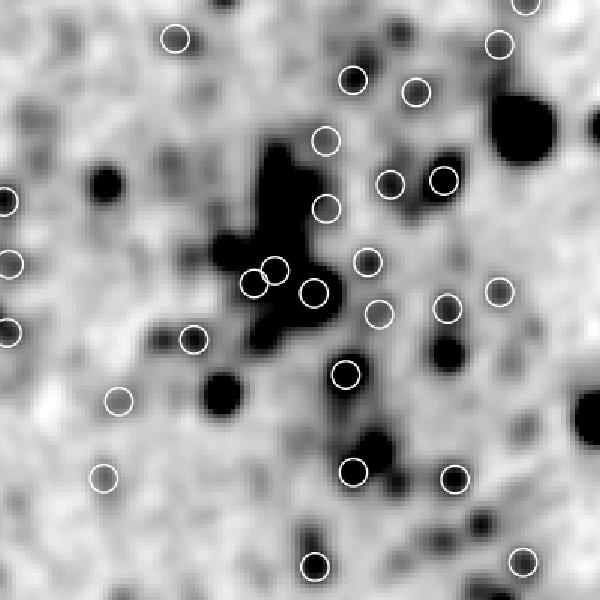}
\hfill
 \includegraphics[width=0.49\linewidth]{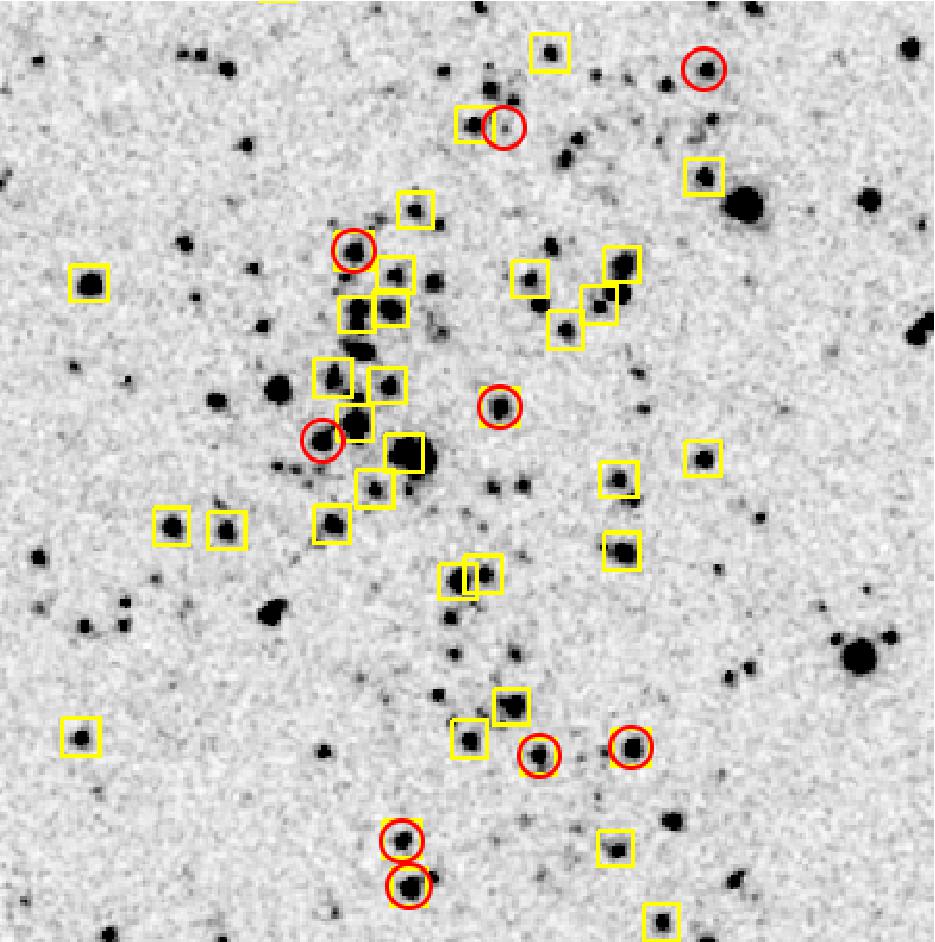}
 \caption{MOO~1210$+$3154, $z = 1.05$, as above; W1 image on the left, $3.6 \mu$m-band image on the right.
 }
 \label{fig:m1210p3153img}
\end{figure*}

\setcounter{figure}{0}
\makeatletter 
\renewcommand{\thefigure}{\@arabic\c@figure l}
\makeatother

\begin{figure*}
\includegraphics[width=0.49\linewidth]{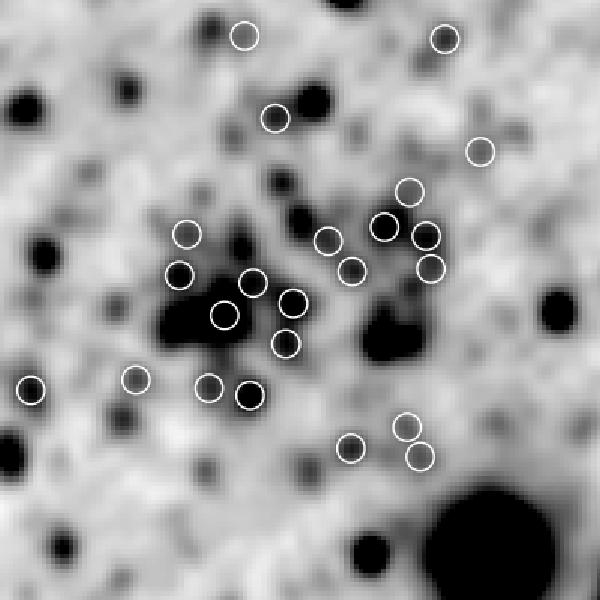}
\hfill
\includegraphics[width=0.49\linewidth]{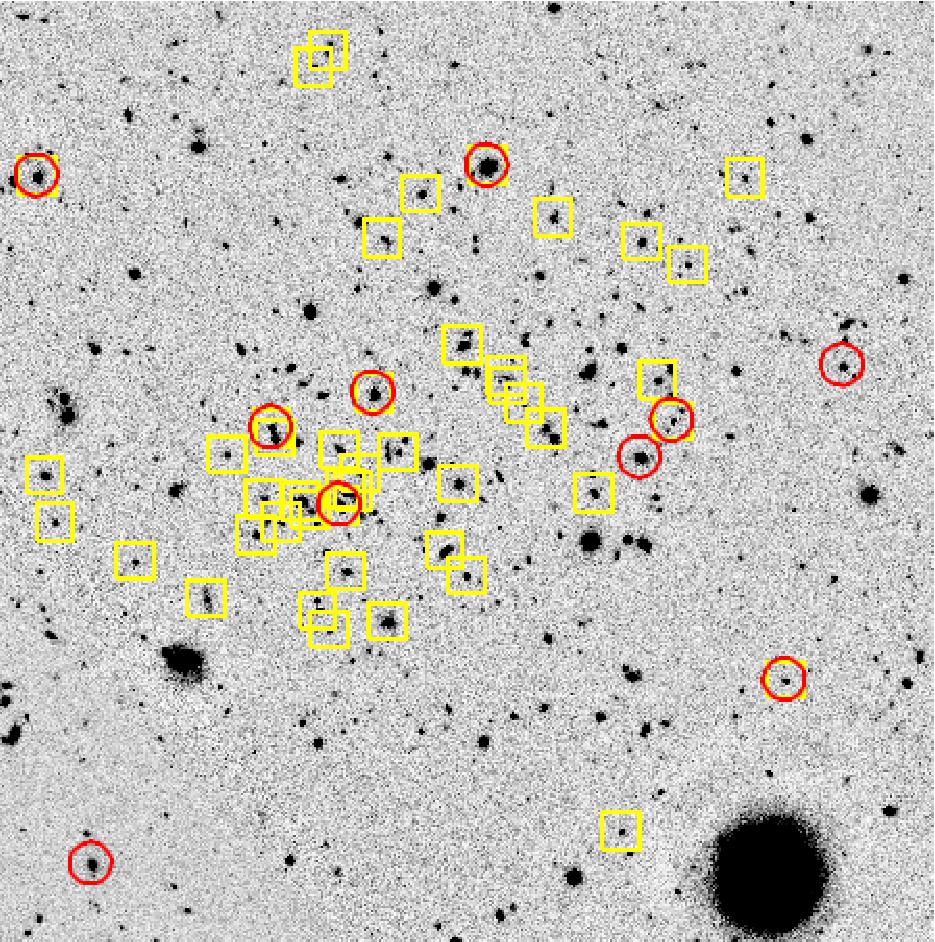}
 \caption{MOO~1319$+$5519, $z = 0.94$, as above; W1 image on the left, $z$-band image on the right. 
 }
 \label{fig:m1319p5519img}
\end{figure*}

\setcounter{figure}{0}
\makeatletter 
\renewcommand{\thefigure}{\@arabic\c@figure m}
\makeatother

\begin{figure*}
\includegraphics[width=0.49\linewidth]{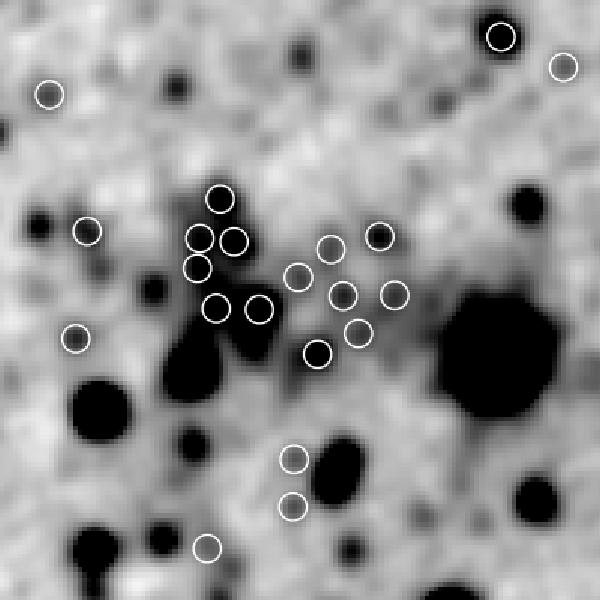}
\hfill
 \includegraphics[width=0.49\linewidth]{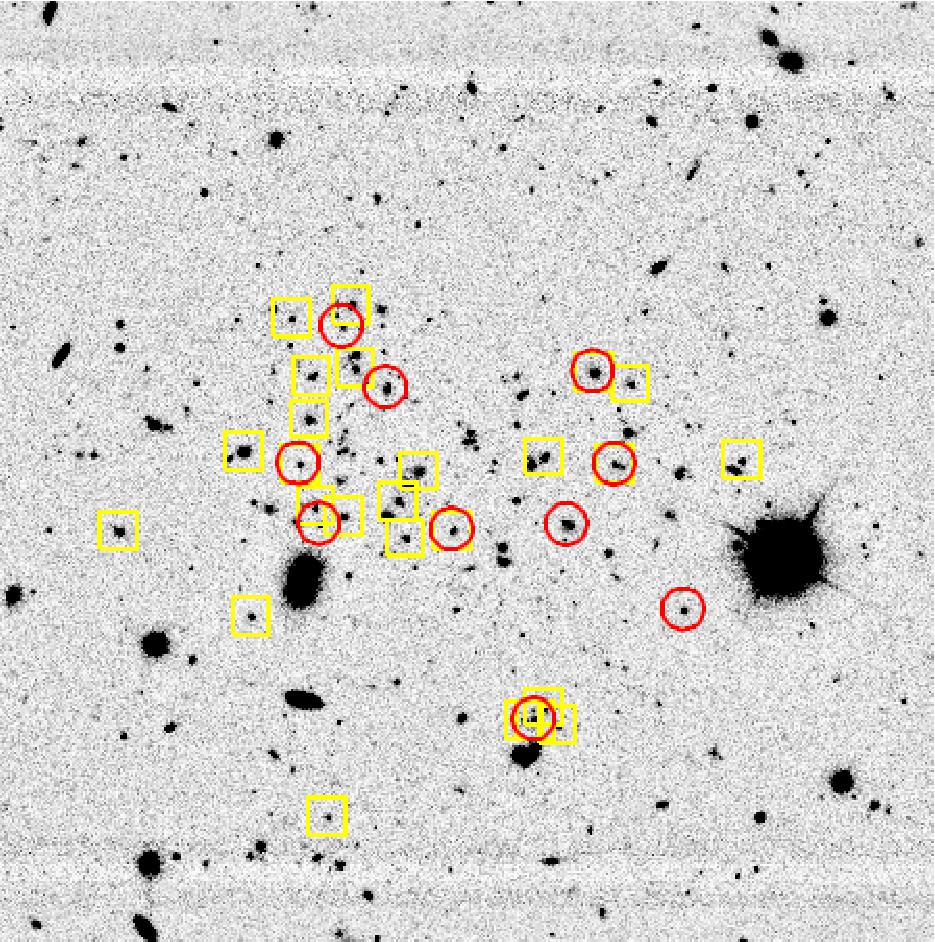}
 \caption{MOO~1335$+$3004, $z = 0.98$, as above; W1 image on the left, $z$-band image on the right. 
 }
 \label{fig:m1335p3004img}
\end{figure*}

\setcounter{figure}{0}
\makeatletter 
\renewcommand{\thefigure}{\@arabic\c@figure n}
\makeatother

\begin{figure*}
\includegraphics[width=0.49\linewidth]{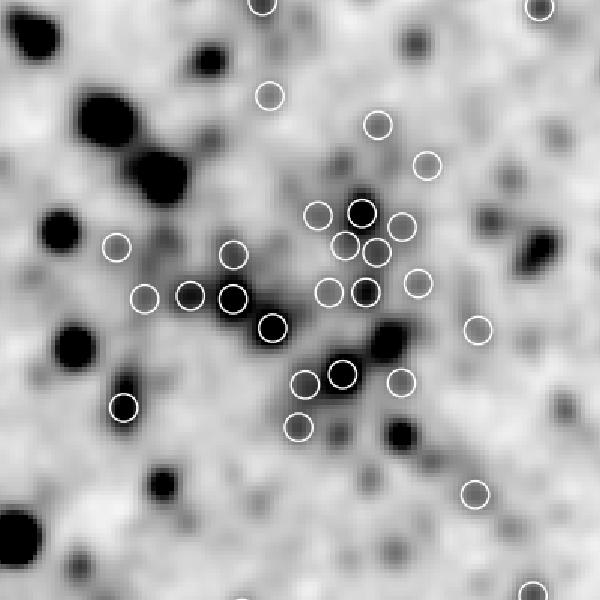}
\hfill
 \includegraphics[width=0.49\linewidth]{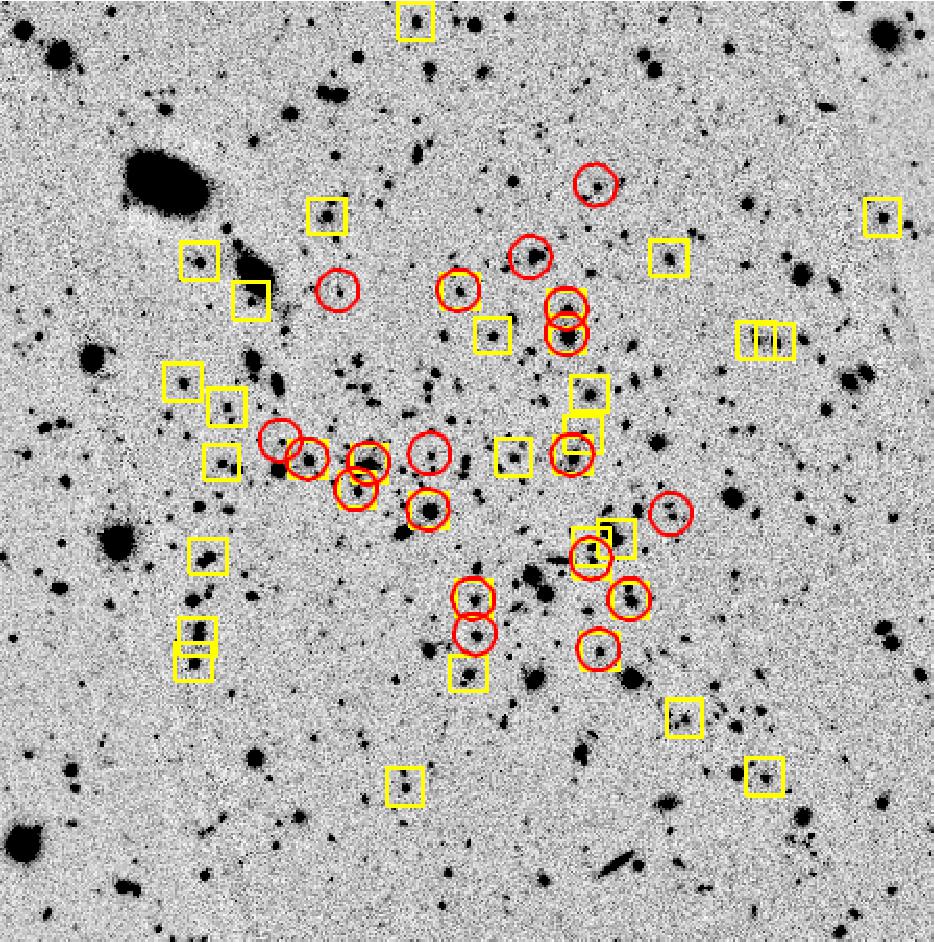}
 \caption{MOO~1514$+$1346, $z = 1.06$, as above; W1 image on the left, $z$-band image on the right.
 }
 \label{fig:m1514p1346img}
\end{figure*}

\setcounter{figure}{0}
\makeatletter 
\renewcommand{\thefigure}{\@arabic\c@figure o}
\makeatother

\begin{figure*}
\includegraphics[width=0.49\linewidth]{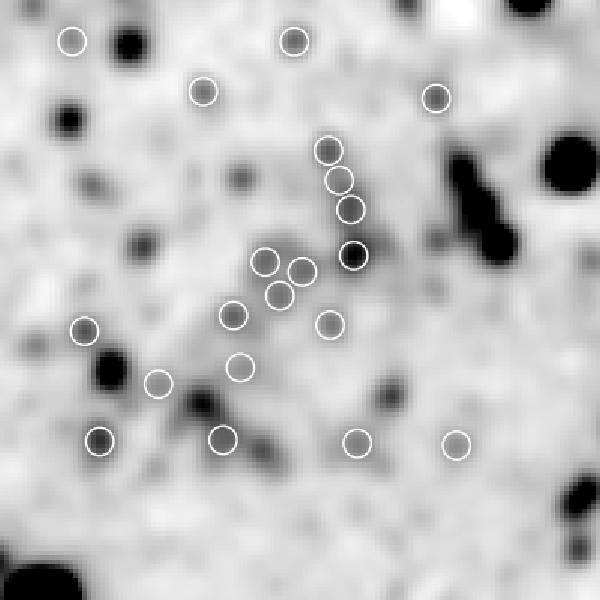}
\hfill
 \includegraphics[width=0.49\linewidth]{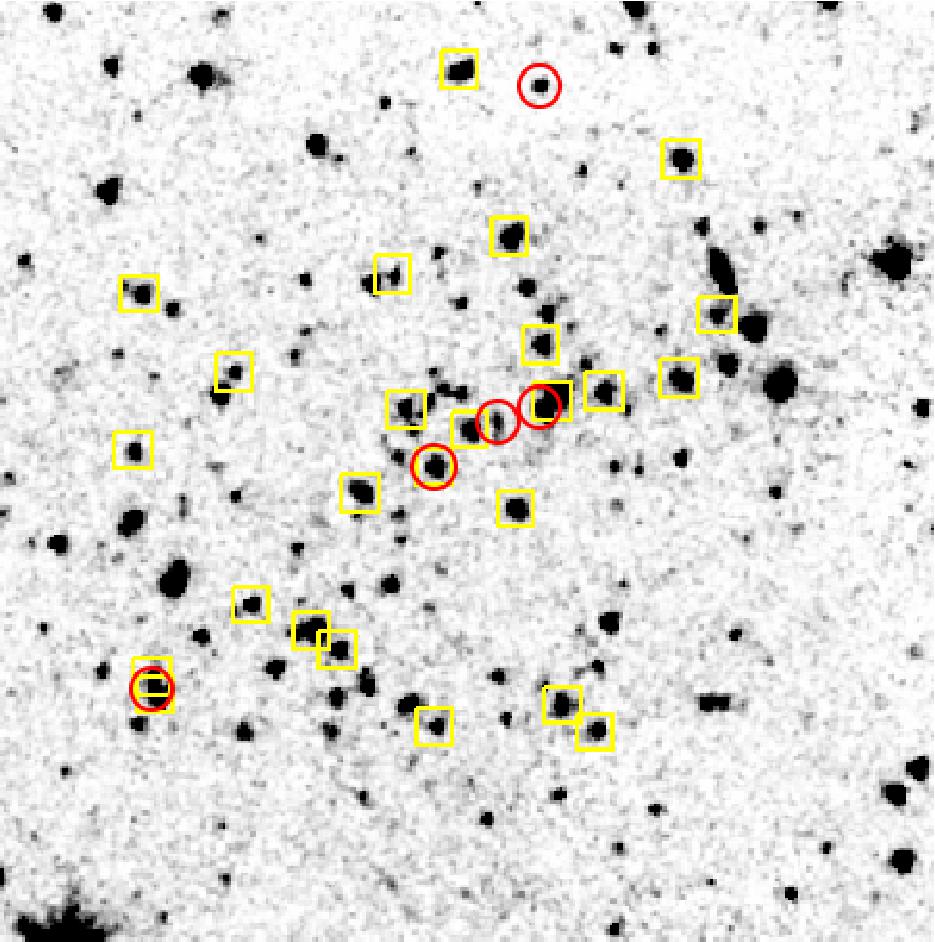}
 \caption{MOO~1625$+$2629, $z = 1.20$, as above; W1 image on the left, $3.6 \mu$m-band image on the right. 
 }
 \label{fig:m1625p2629img}
\end{figure*}

\clearpage

\setcounter{figure}{0}
\makeatletter 
\renewcommand{\thefigure}{\@arabic\c@figure p}
\makeatother

\begin{figure*}
 \includegraphics[width=0.49\linewidth]{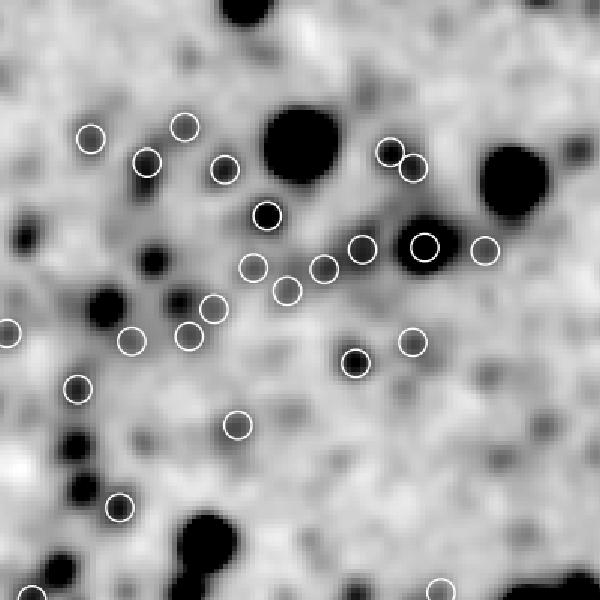}
 \hfill
 \includegraphics[width=0.49\linewidth]{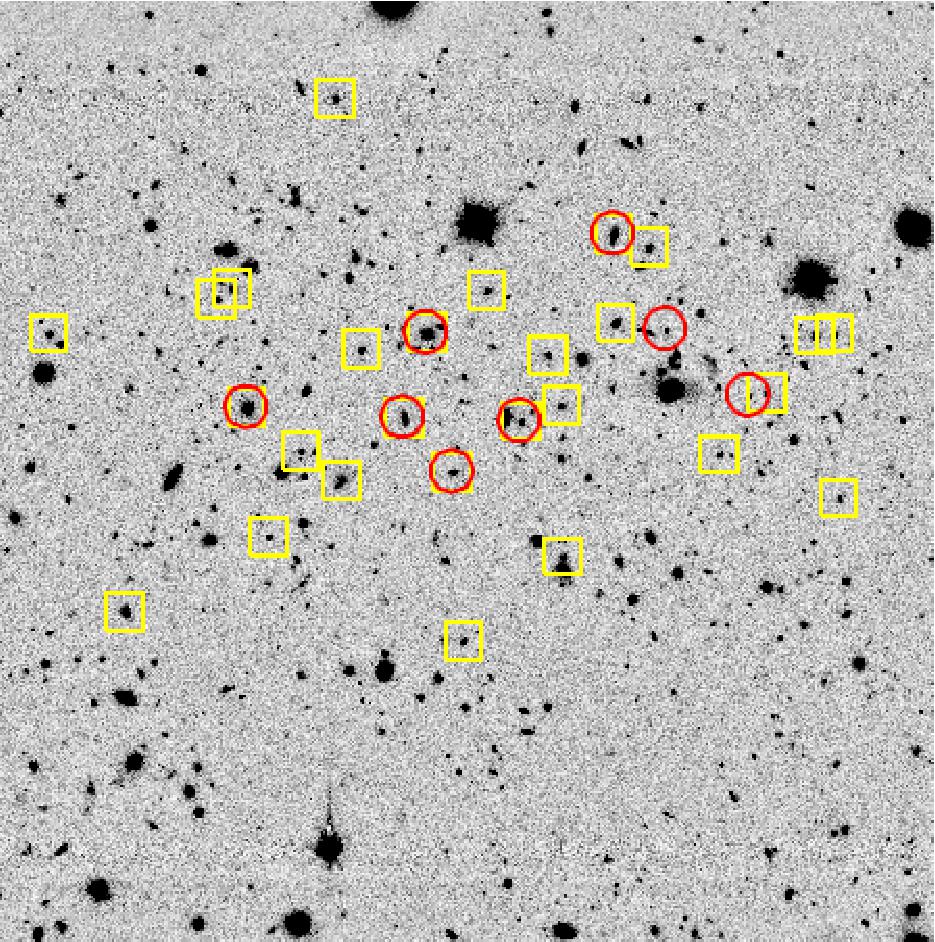}
 \caption{MOO~2205$-$0917, $z = 0.93$, as above; W1 image on the left, $z$-band image on the right.
 }
 \label{fig:m2205m0917img}
\end{figure*}

\setcounter{figure}{0}
\makeatletter 
\renewcommand{\thefigure}{\@arabic\c@figure q}
\makeatother

\begin{figure*}
 \includegraphics[width=0.49\linewidth]{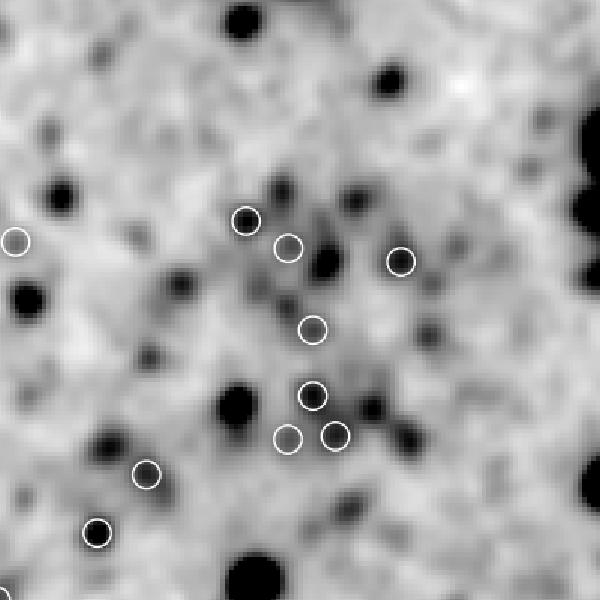}
 \hfill
 \includegraphics[width=0.49\linewidth]{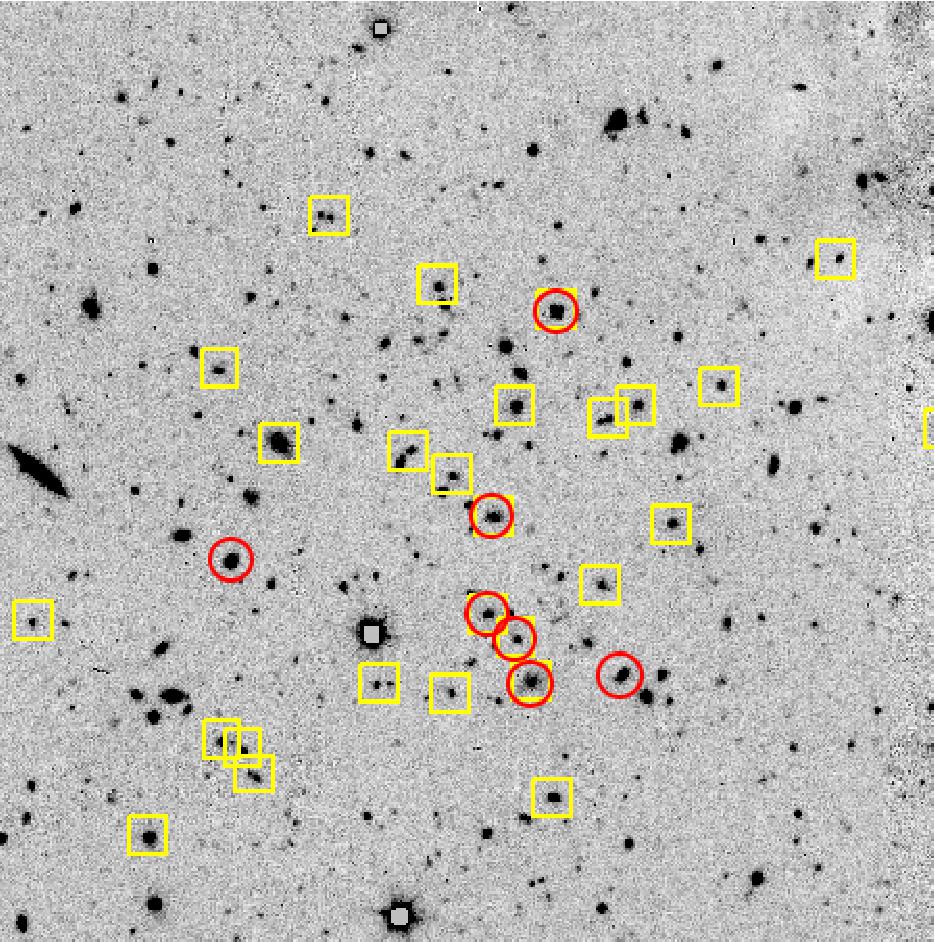}
 \caption{MOO~2320$-$0620, $z = 0.92$, as above; W1 image on the left, $i$-band image on the right.
 }
 \label{fig:m2320m0620img}
\end{figure*}

\setcounter{figure}{0}
\makeatletter 
\renewcommand{\thefigure}{\@arabic\c@figure r}
\makeatother

\begin{figure*}
 \includegraphics[width=0.49\linewidth]{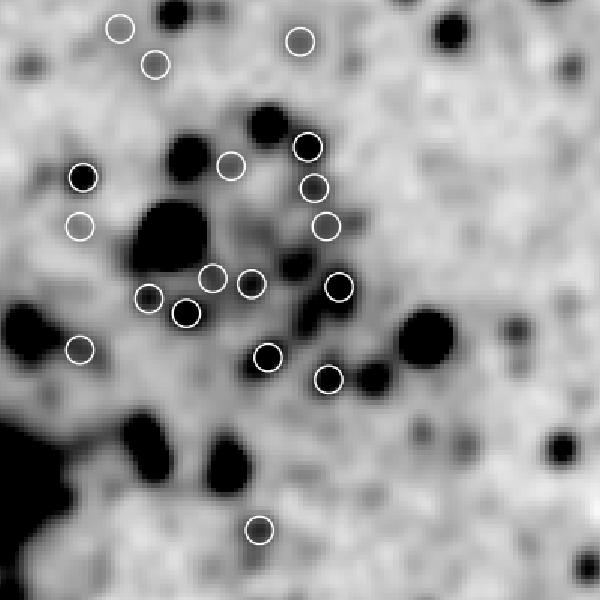}
 \hfill
 \includegraphics[width=0.49\linewidth]{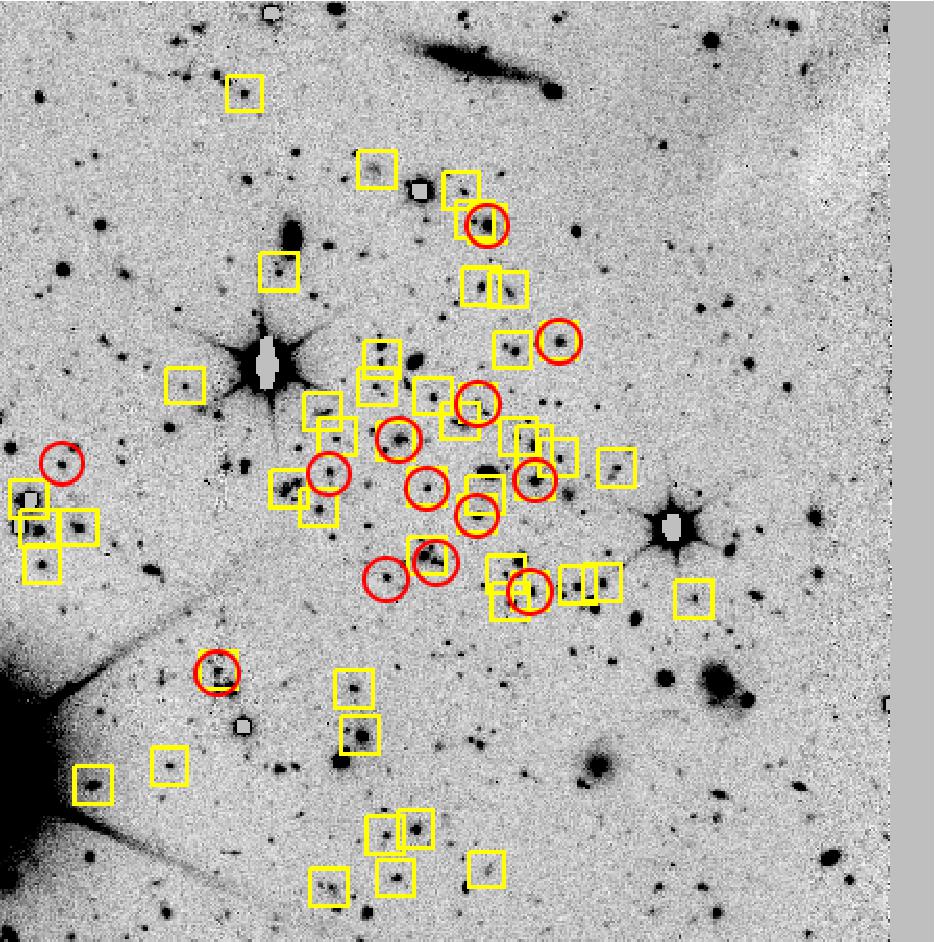}
 \caption{MOO~2348$+$0846, $z = 0.89$, as above; W1 image on the left, $i$-band image on the right.
 }
 \label{fig:m2348p0846img}
\end{figure*}

\setcounter{figure}{0}
\makeatletter 
\renewcommand{\thefigure}{\@arabic\c@figure s}
\makeatother

\begin{figure*}
 \includegraphics[width=0.49\linewidth]{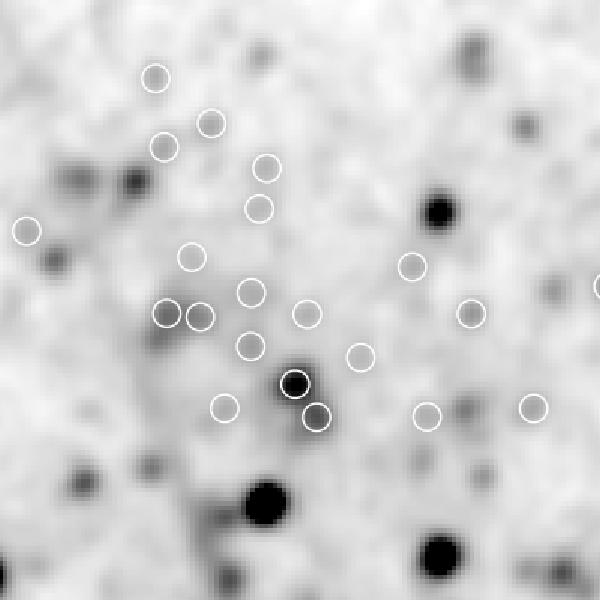}
 \hfill
 \includegraphics[width=0.49\linewidth]{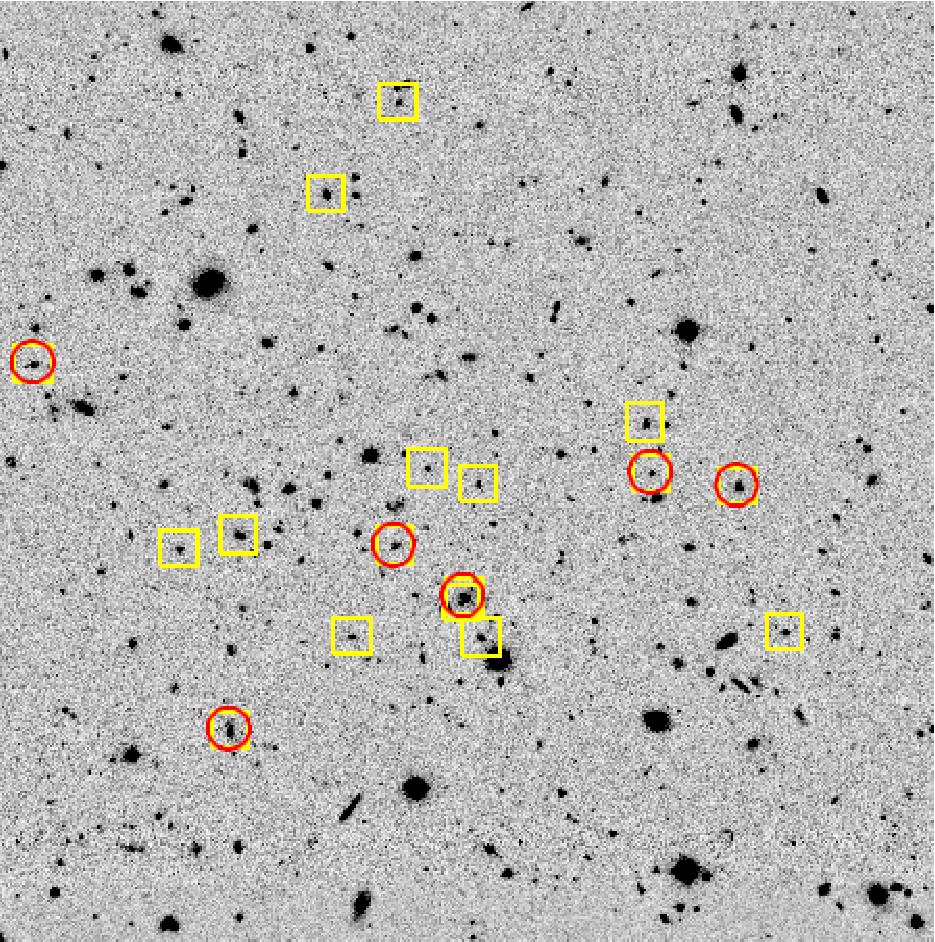}
 \caption{MOO~2355$+$1030, $z = 1.27$, as above; W1 image on the left, $z$-band image on the right.
 }
 \label{fig:m2355p1030img}
\end{figure*}

\begin{deluxetable}{cllllc}
\tabletypesize{\scriptsize}
\tablewidth{0pt}
\tablecaption{Redshifts of Spectroscopic Members}
\tablehead{\colhead{Object\tablenotemark{1}} & \colhead{$\alpha$ (J2000)} & \colhead{$\delta$ (J2000)}  & \colhead{z} & \colhead{Q} &\colhead{Type}   }
\startdata
0012+1602.001213.6+160204 & 00:12:13.699 &   16:02:04.7 &  0.946 &  B & E \\ 
0012+1602.001218.7+160246  & 00:12:18.742 &   16:02:46.0 &  0.946 &  A & A \\
0012+1602.001211.2+160151 & 00:12:11.249 &   16:01:51.3 &  0.940 &  B & A \\ 
0012+1602.001209.9+160216 & 00:12:09.983 &   16:02:16.4 &  0.950 &  A & A \\ 
0012+1602.001208.6+160238 & 00:12:08.601 &   16:02:38.2 &  0.949 &  B & A \\ 
0012+1602.001214.7+160231  & 00:12:14.799 &   16:02:31.6 &  0.938 &  A & E \\ 
0012+1602.001218.7+160219 & 00:12:18.732 &   16:02:19.1 &  0.941 &  B & A \\ 
0012+1602.001212.5+160214 & 00:12:12.507 &   16:02:14.4 &  0.950 &  B & A \\ 
0012+1602.001206.8+160217 & 00:12:06.854 &   16:02:17.2 &  0.952 &  B & A \\ 
0012+1602.001217.1+160227  & 00:12:17.166 &   16:02:27.6 &  0.943 &  A & A \\ 
0012+1602.001214.2+160220  & 00:12:14.223 &   16:02:20.5 &  0.941 &  A & E \\ 
0012+1602.001216.3+160241  & 00:12:16.302 &   16:02:41.2 &  0.938 &  A & E \\ 
0012+1602.001218.7+160314  & 00:12:18.799 &   16:03:14.7 &  0.932 &  A & E \\ 
0012+1602.001213.4+160234  & 00:12:13.415 &   16:02:34.1 &  0.952 &  B & A \\ 
0012+1602.001216.1+160229  & 00:12:16.104 &   16:02:29.3 &  0.952 &  A & A \\ 
0012+1602.001223.6+155749 & 00:12:23.627 &   15:57:49.7 &  0.948 &  A & E \\ 
0012+1602.001229.7+155802 & 00:12:29.705 &   15:58:02.4 &  0.947 &  A & E \\ 
0012+1602.001215.4+160236  & 00:12:15.437 &   16:02:36.8 &  0.942 &  A & E \\ 
0012+1602.001213.4+160155 & 00:12:13.403 &   16:01:55.7 &  0.940 &  B & A \\ 
0012+1602.001222.7+160126  & 00:12:22.798 &   16:01:26.1 &  0.930 &  B & A \\ 
0012+1602.001214.0+160033 & 00:12:14.099 &   16:00:33.7 &  0.948 &  A & A \\ 
0012+1602.001214.4+160229  & 00:12:14.418 &   16:02:29.1 &  0.942 &  B & A \\ 
0012+1602.001215.5+160150 & 00:12:15.516 &   16:01:50.0 &  0.944 &  B & A \\ 
\enddata
\tablecomments{Redshifts of individual galaxies.  This is a partial listing of the member galaxies in only one cluster; redshifts of individual galaxies for all the clusters will be available electronically.  The Q column describes the quality of the redshift using the following codes:  A = high degree of certainty based on multiple features; B = reasonable certainty based on at least two features.  In the Type column, E means the optical spectrum has an emission line (nearly always [O II]$\lambda$3727), and A means it is an absorption line (Ca H$+$K, D4000) spectrum. 
}
\tablenotetext{1}{The object ID is created from the cluster MOO ID followed by the RA and Dec of the object.}
\label{redshifts}
\end{deluxetable}

\setcounter{figure}{1}
\makeatletter 
\renewcommand{\thefigure}{\@arabic\c@figure}
\makeatother

\begin{figure}
\setlength\fboxsep{-10.0pt}
\setlength\fboxrule{-11.0pt}
\fbox{ 
  \includegraphics[width=0.40\textwidth]{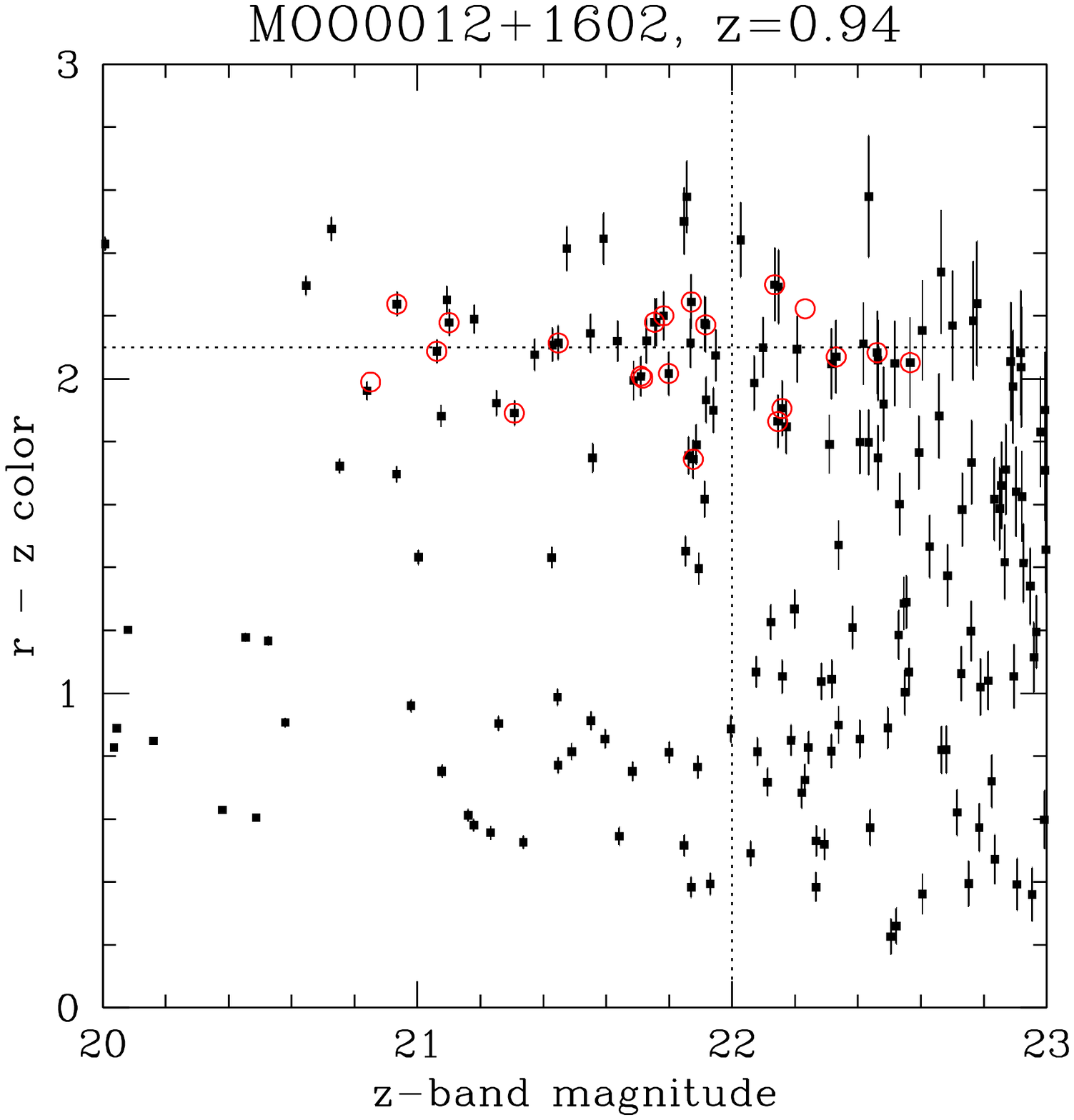}
}
\fbox{
  \includegraphics[width=0.40\textwidth]{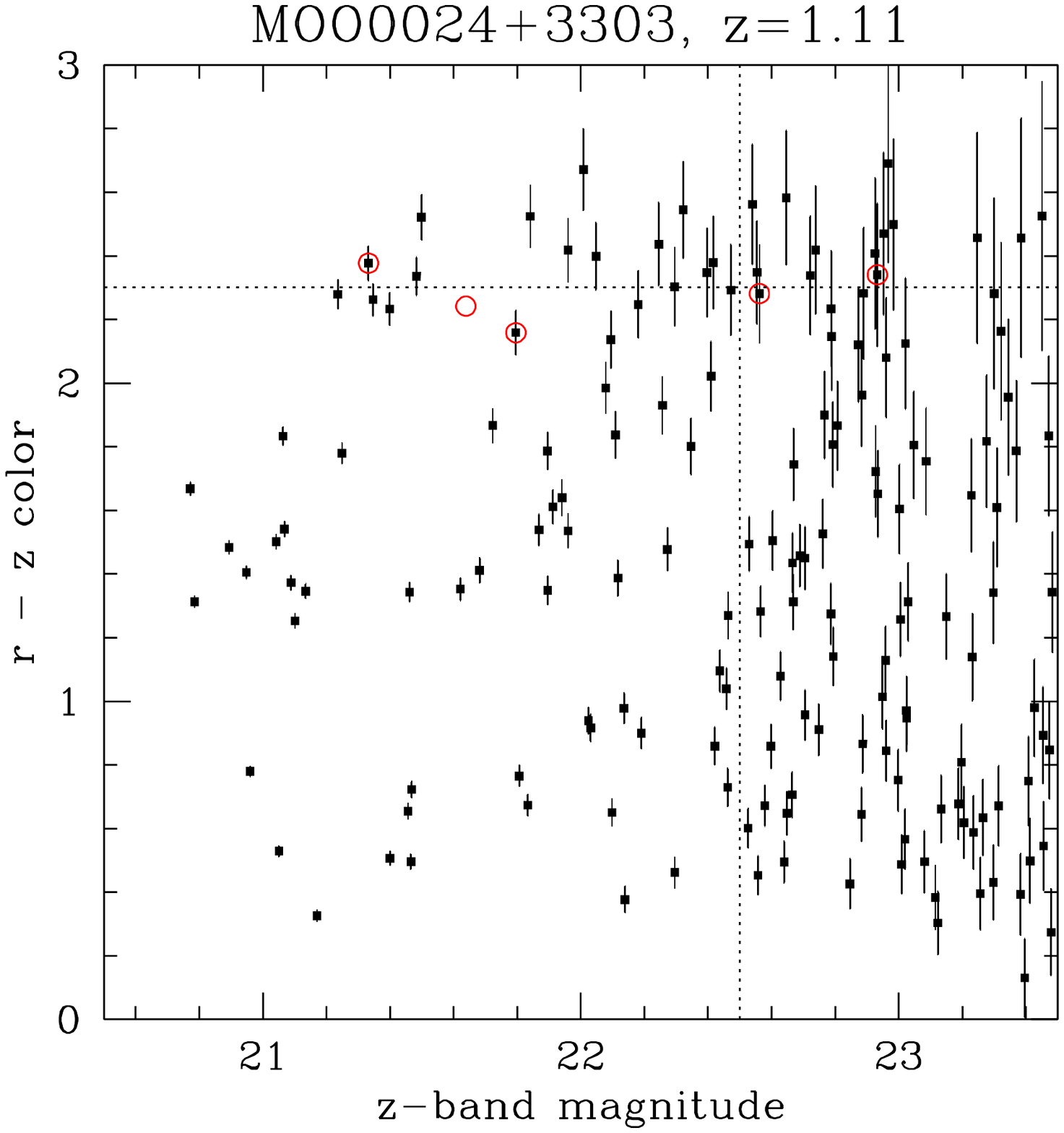}
}
\fbox{
  \includegraphics[width=0.40\textwidth]{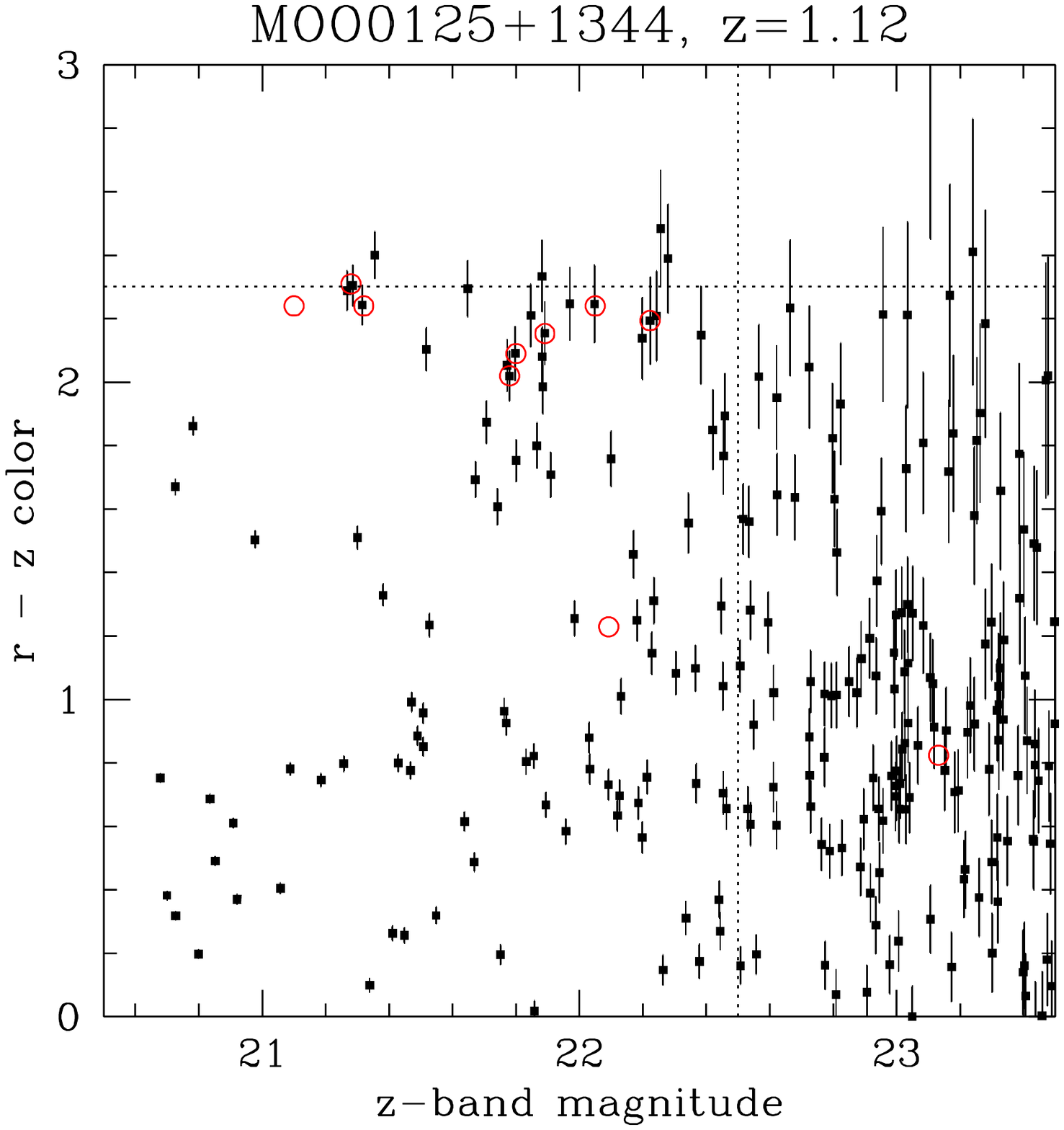}
}
\fbox{
  \includegraphics[width=0.40\textwidth]{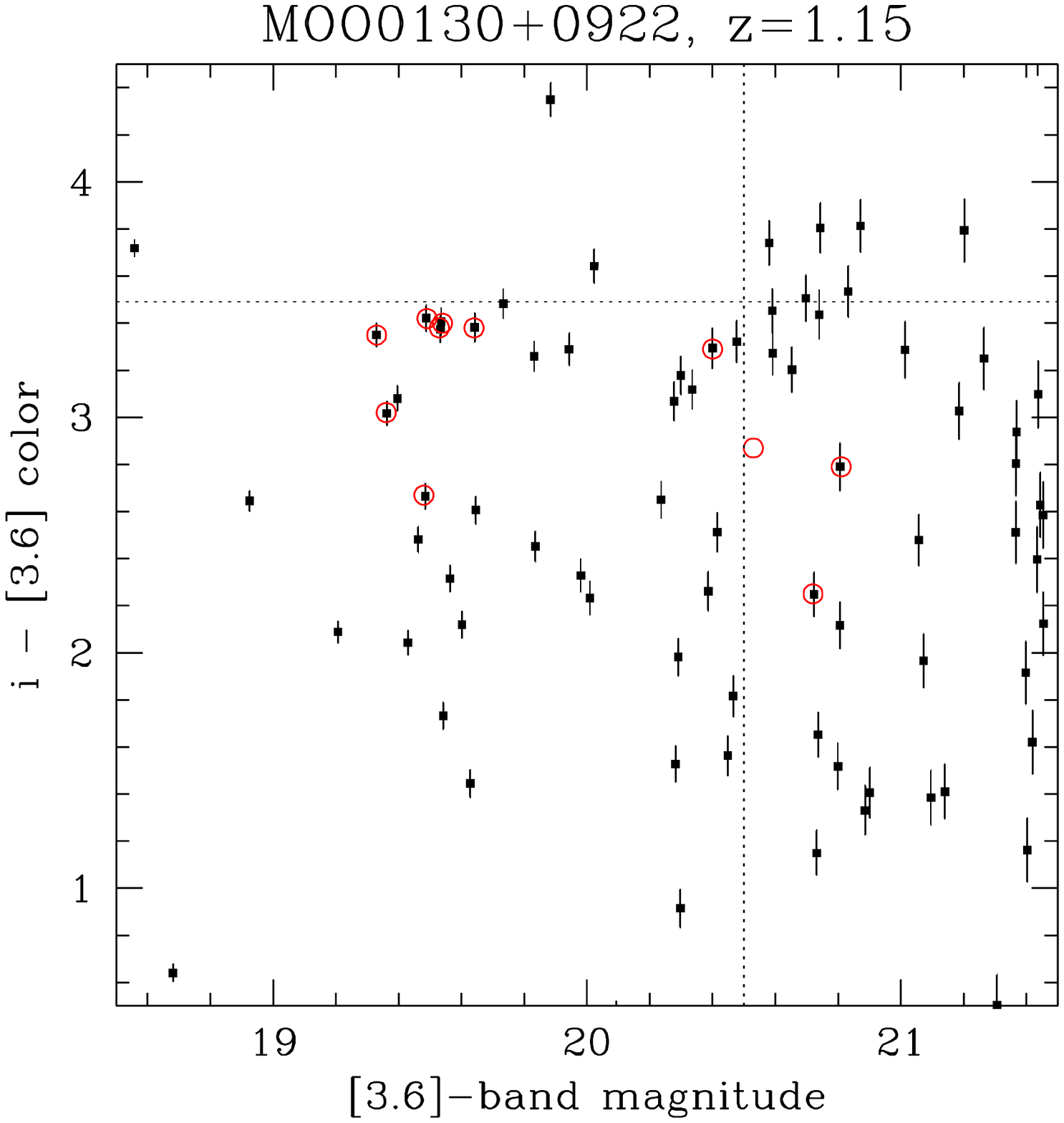}
  }
\fbox{
  \includegraphics[width=0.40\textwidth]{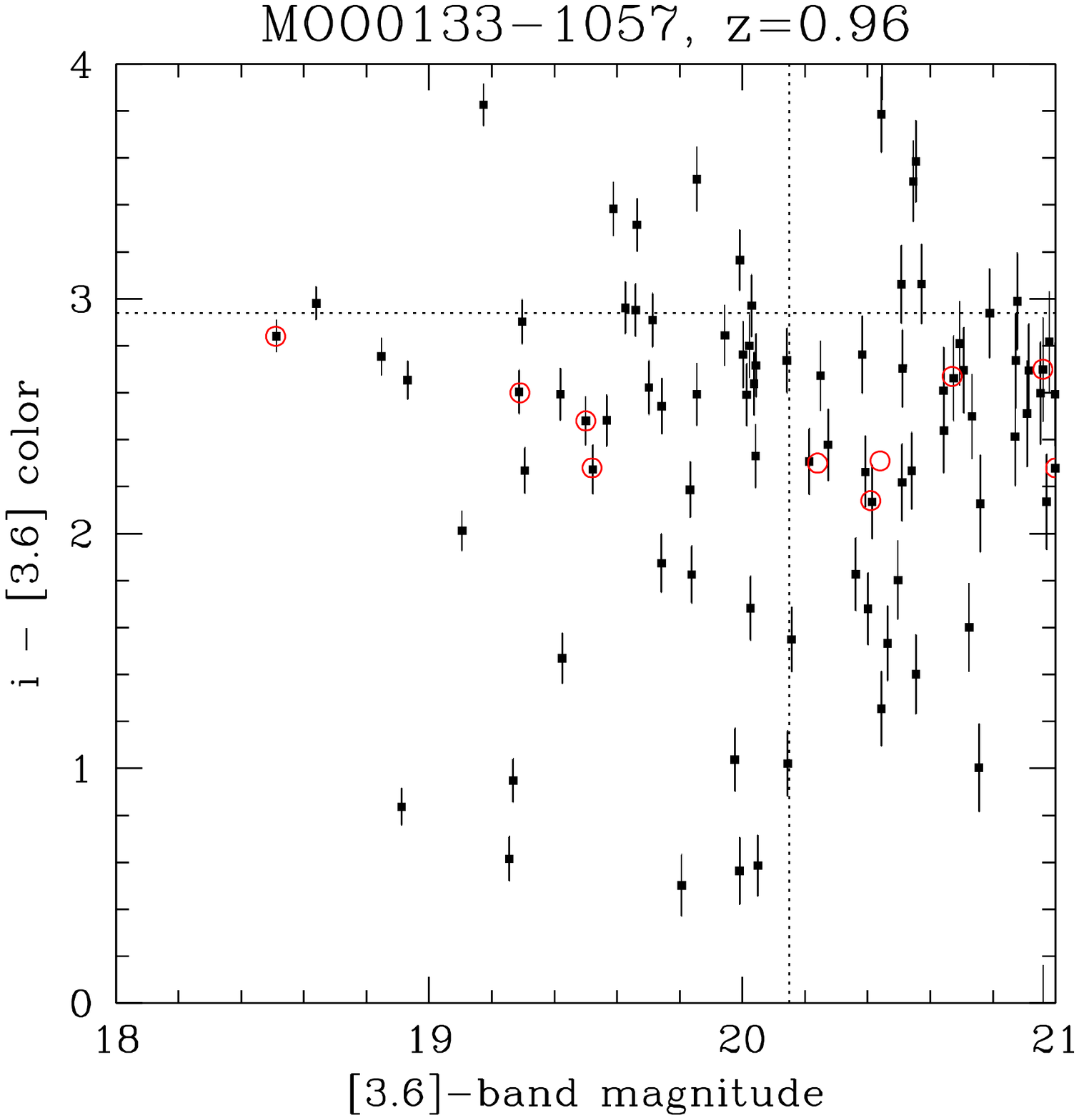}
  }
\fbox{
  \includegraphics[width=0.40\textwidth]{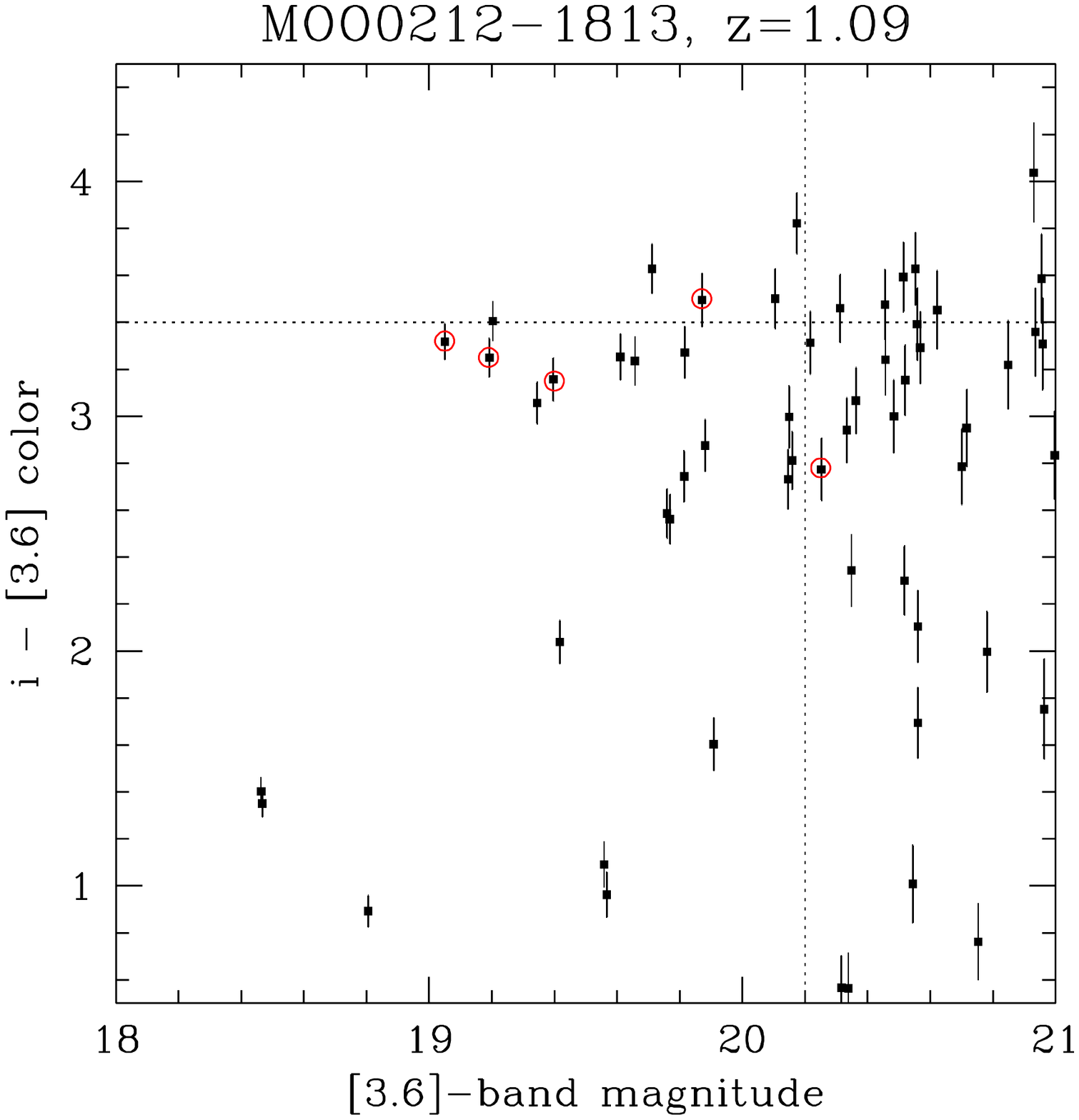}
  }
  \centering
  \caption{Color-magnitude diagrams covering a circular area with a diameter of 3 arcmin ($\sim 1.4$~Mpc).  Uncertainties of one $\sigma$ in the colors are shown.  The color and magnitude of an L$^*$ model galaxy (described in the text) at the cluster redshift are represented by the horizontal and vertical dashed lines.   Spectroscopic members are marked by red circles; members outside the 3 arcmin diameter area are represented by only a red circle.  Some members lie outside of the follow-up imaging area and so are not shown in the CMD.  
}
\label{fig:cmd}
\end{figure}

\setcounter{figure}{1}

\begin{figure}
\setlength\fboxsep{-10.0pt}
\setlength\fboxrule{0.0pt}
\fbox{  
    \includegraphics[width=0.45\textwidth]{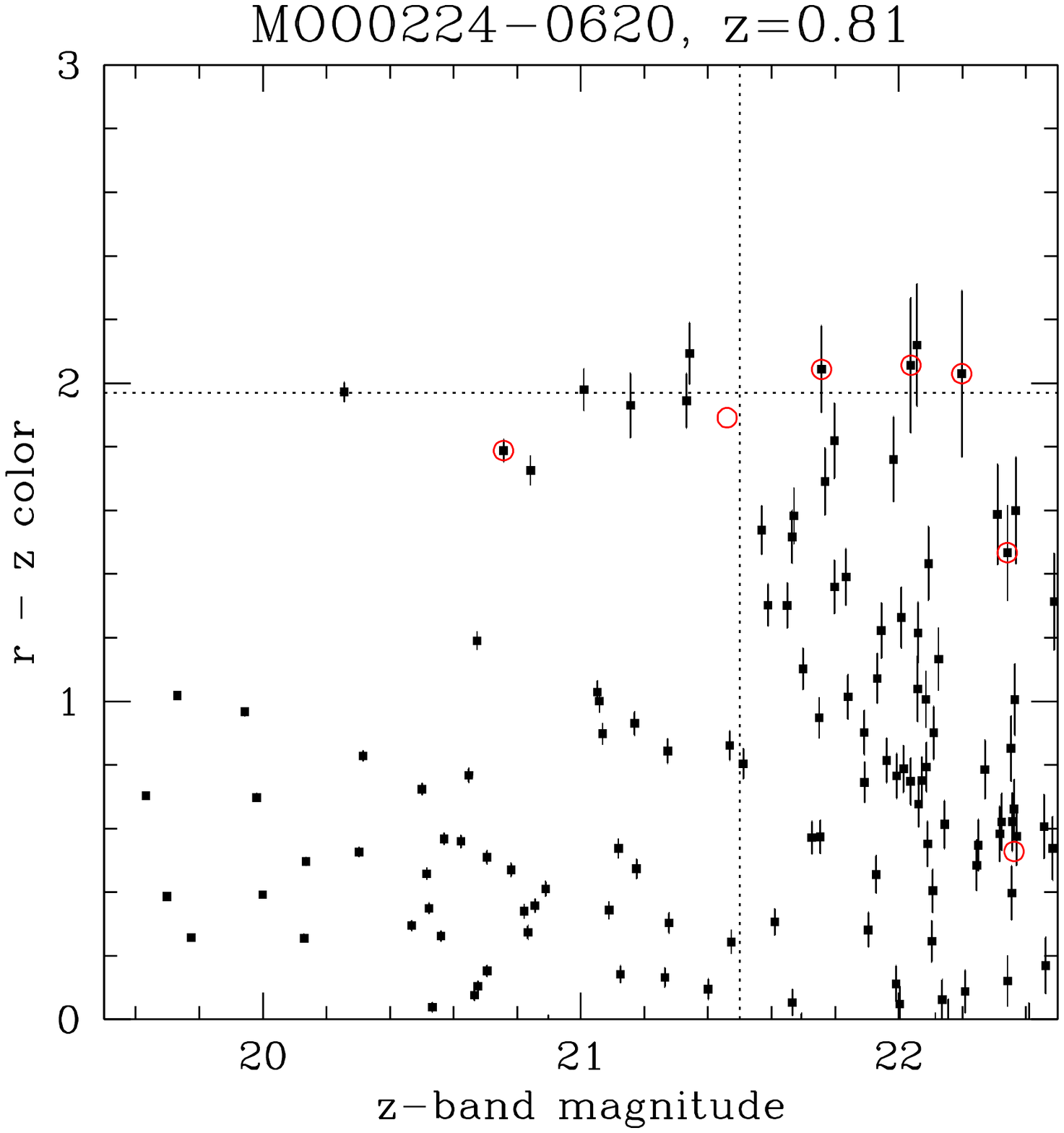}
    }
\vspace{-2.5cm}
\fbox{
  \includegraphics[width=0.45\textwidth]{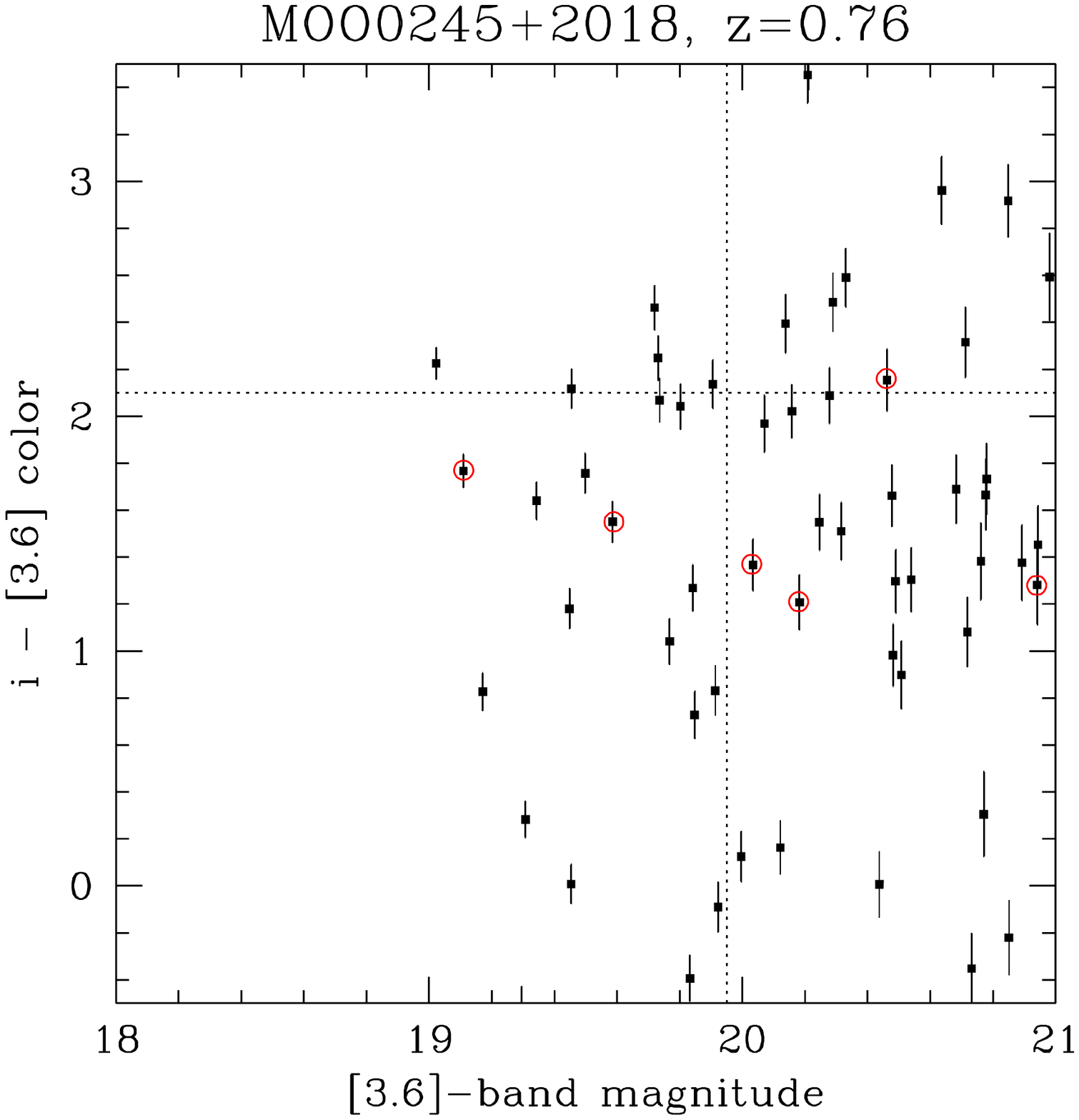}
  }
\fbox{  
    \includegraphics[width=0.45\textwidth]{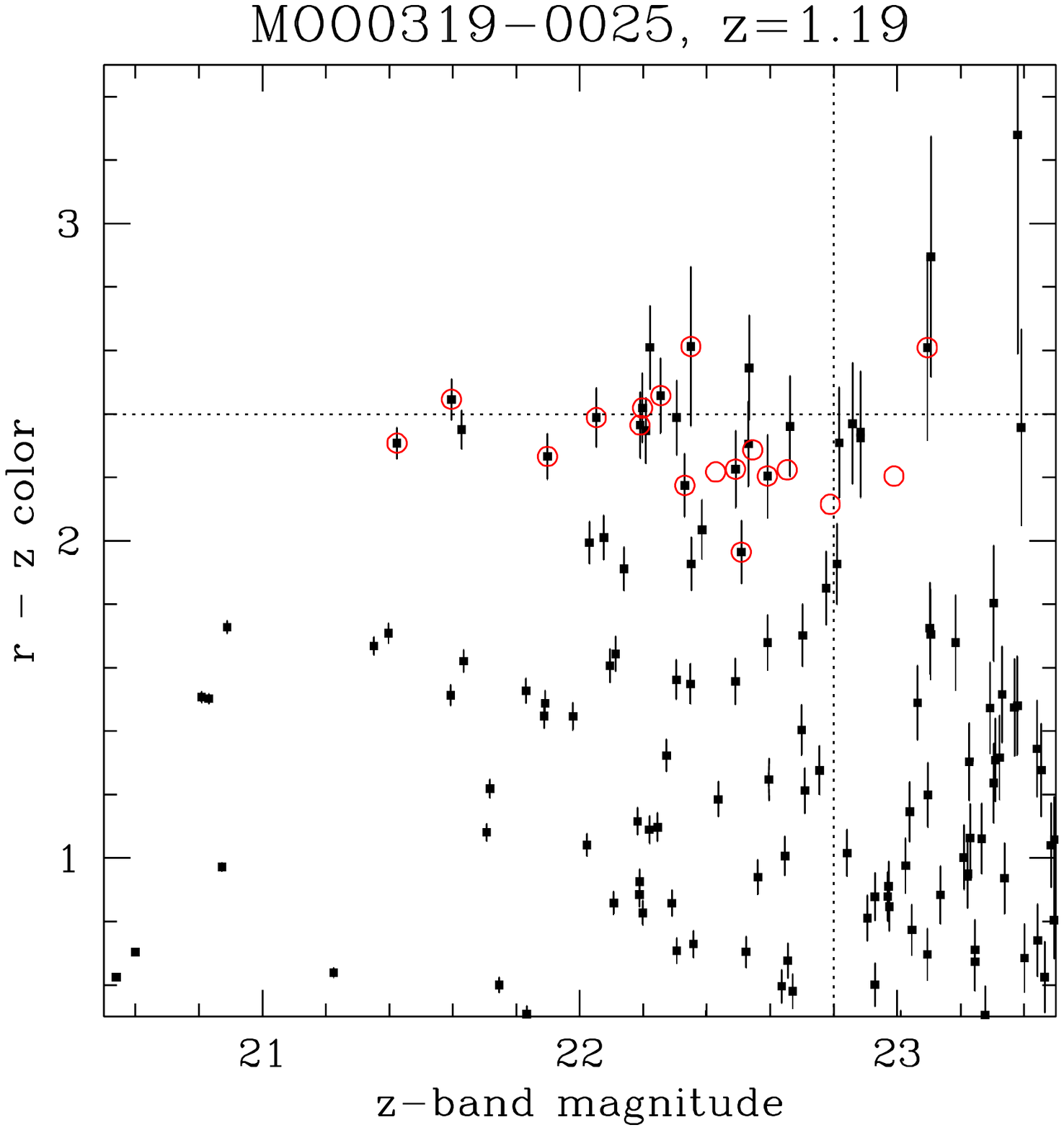}
    }
\vspace{-2.5cm}
 \fbox{
  \includegraphics[width=0.45\textwidth]{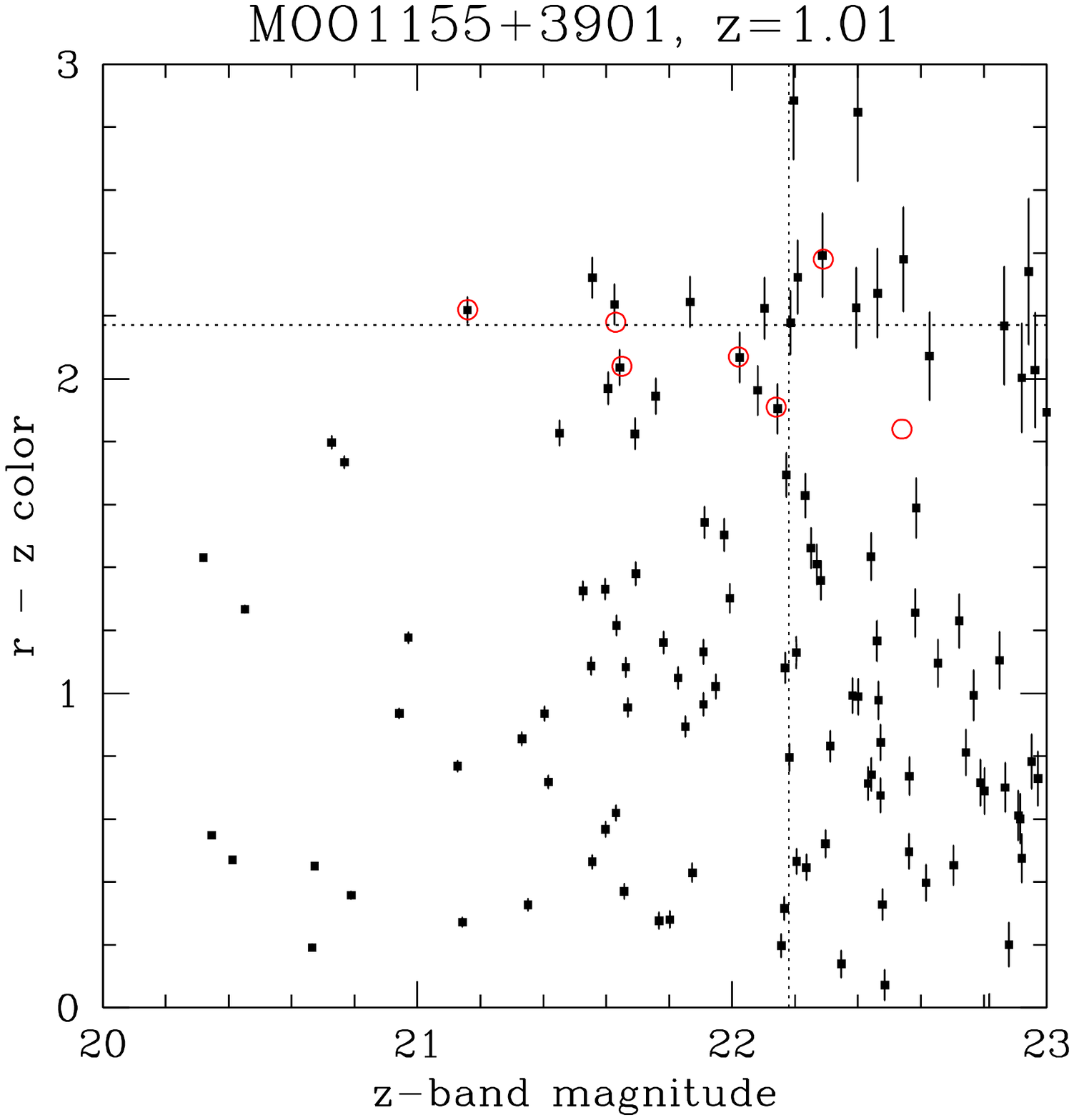}
  }
\fbox{
  \includegraphics[width=0.45\textwidth]{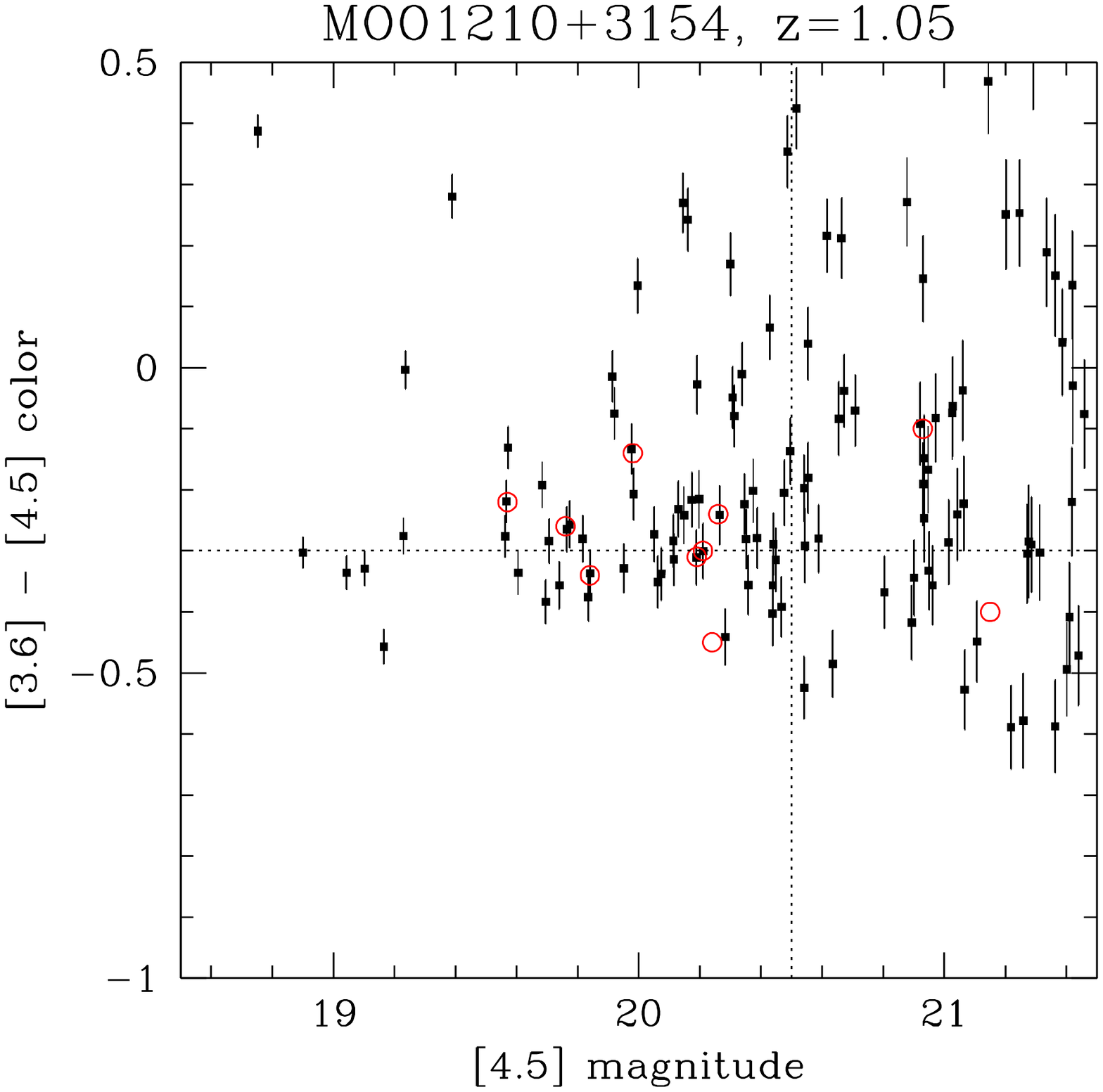}
  }
\fbox{  
    \includegraphics[width=0.45\textwidth]{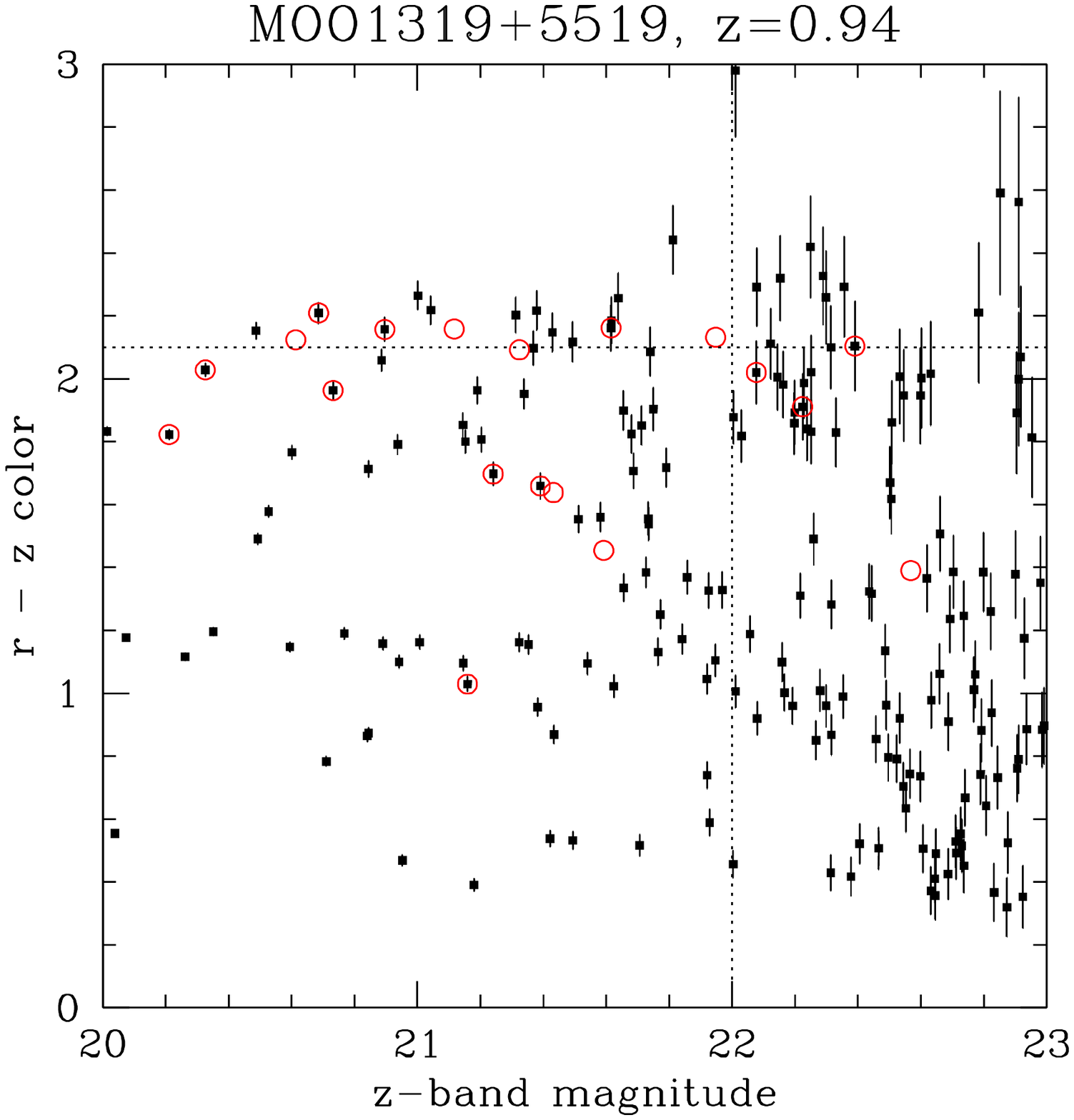}
    }
 \centering
 \vspace{-1.0cm}
 \caption{Continued.}
\end{figure}

\setcounter{figure}{1}

\begin{figure}
\setlength\fboxsep{-10.0pt}
\setlength\fboxrule{0.0pt}
\fbox{
  \includegraphics[width=0.45\textwidth]{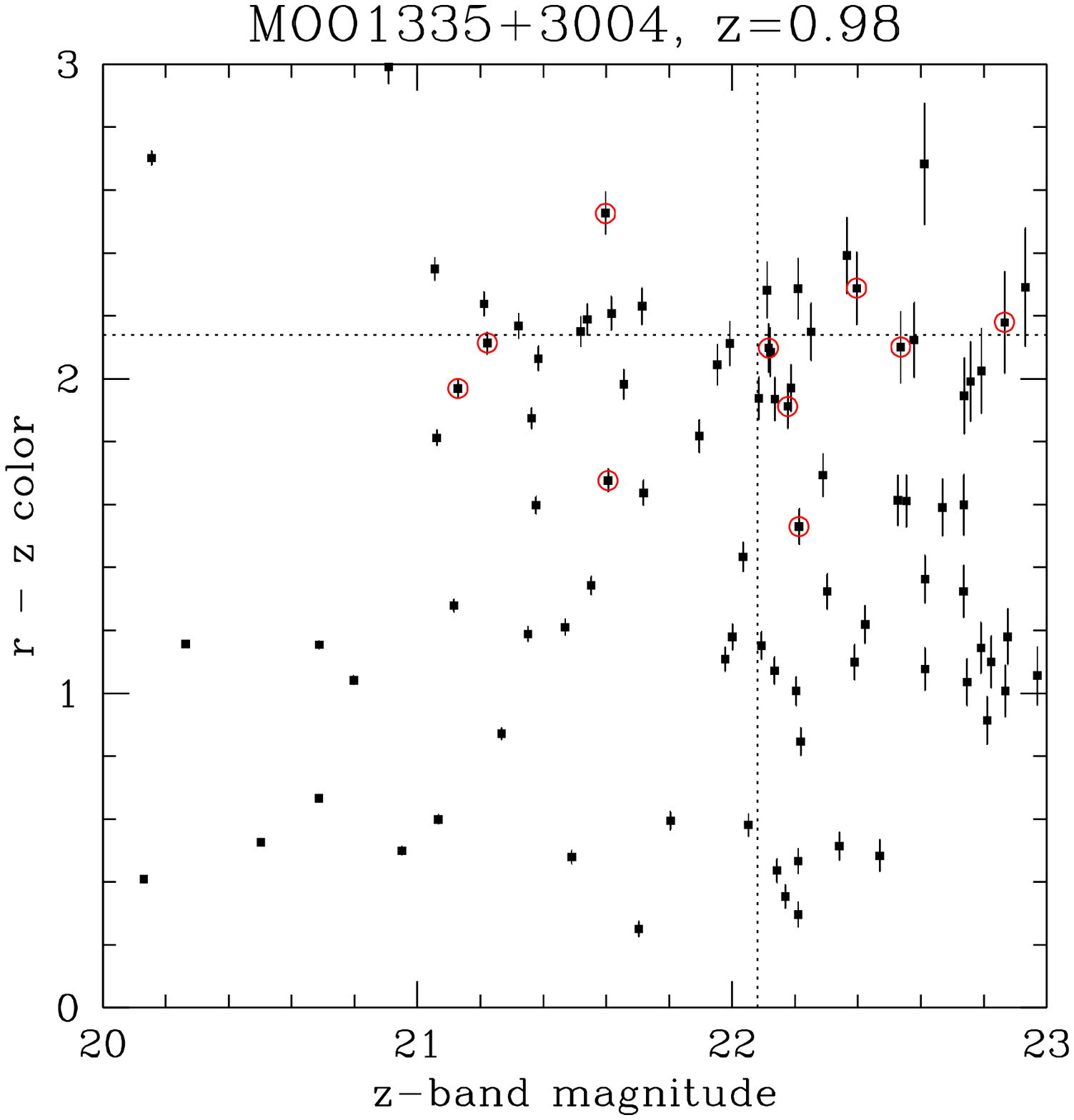}
  }
\vspace{-2.5cm}
\fbox{
  \includegraphics[width=0.45\textwidth]{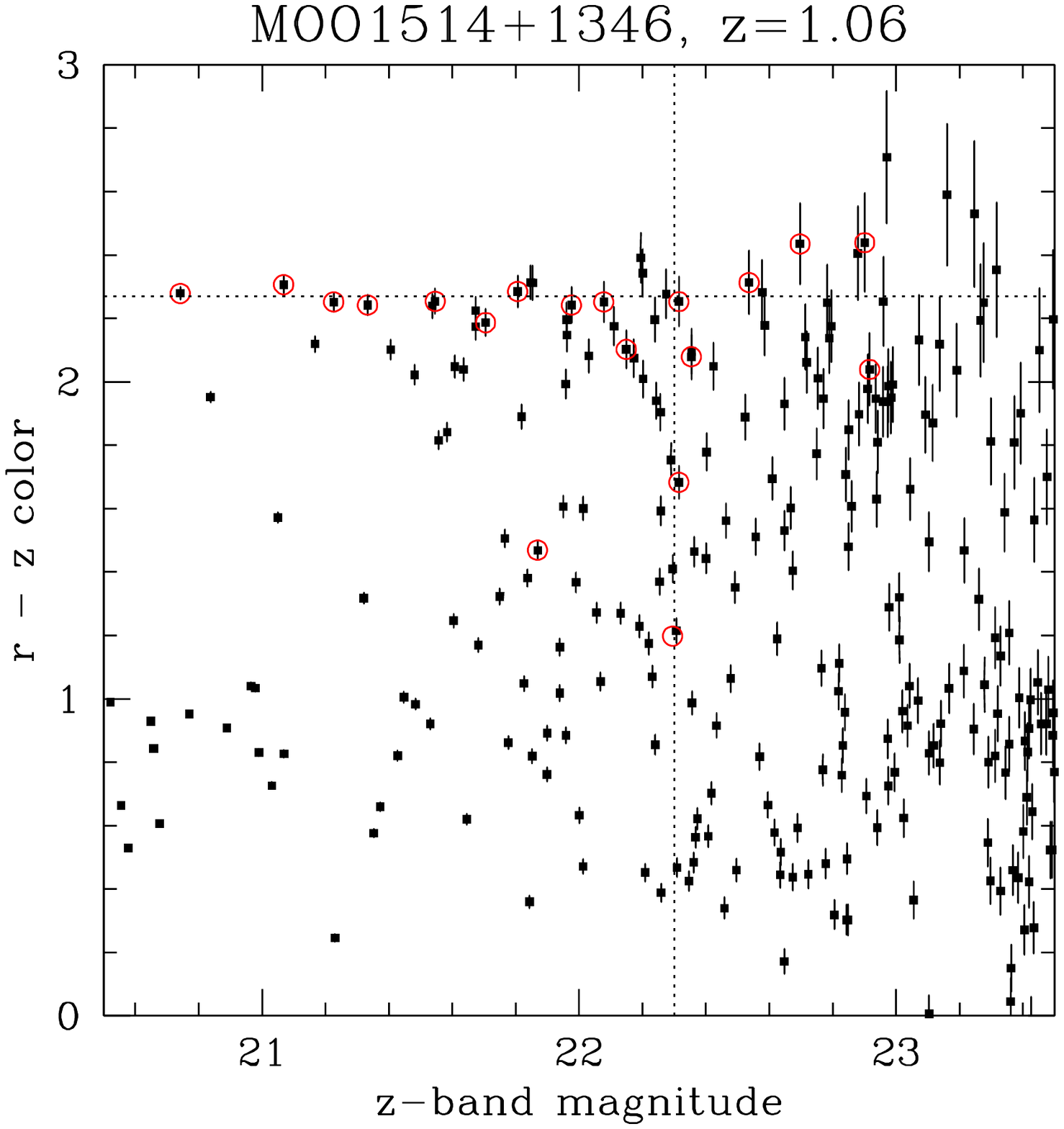}
  }
\fbox{  
\includegraphics[width=0.45\textwidth]{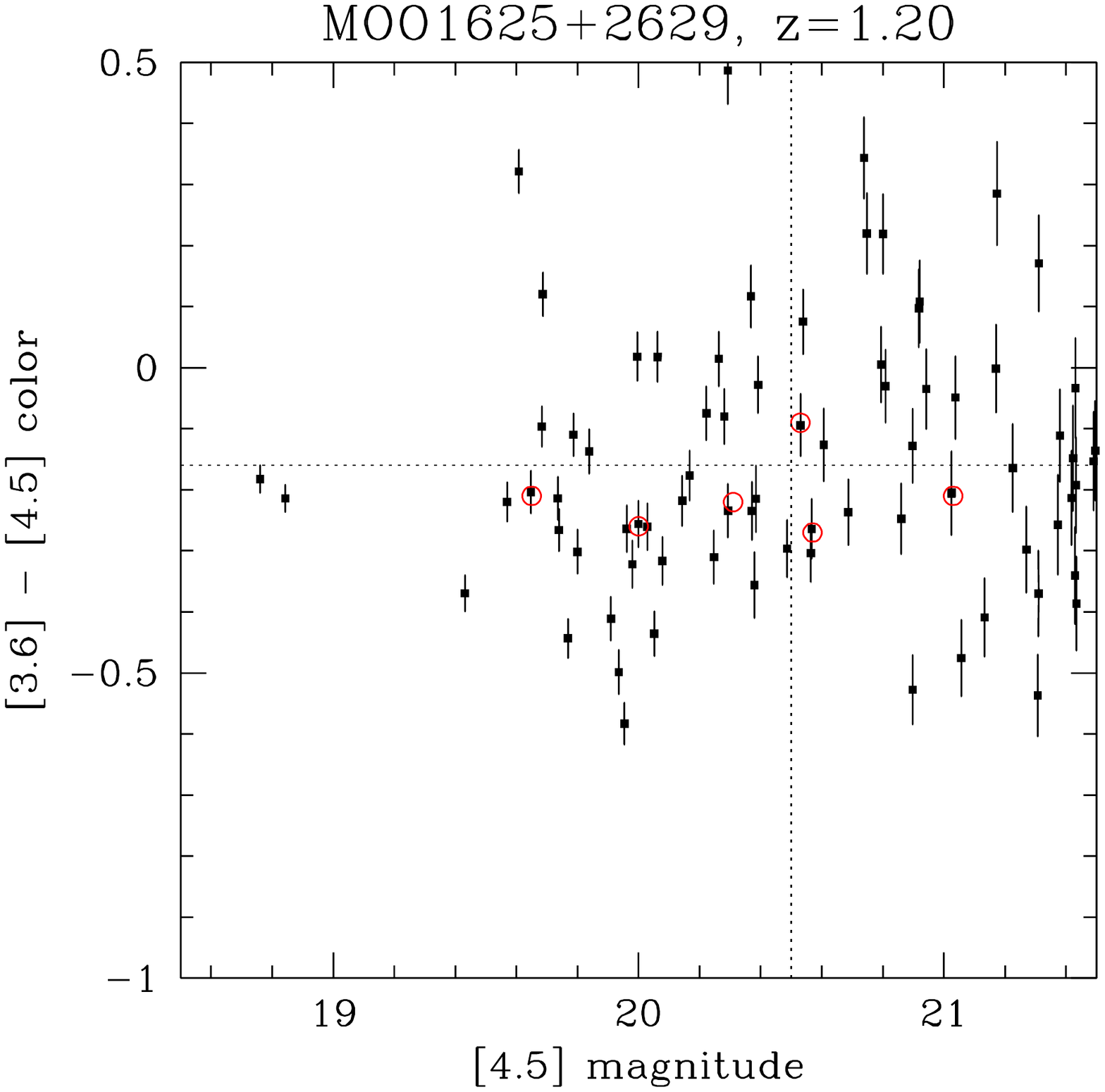}
}
\vspace{-2.5cm}
\fbox{
  \includegraphics[width=0.45\textwidth]{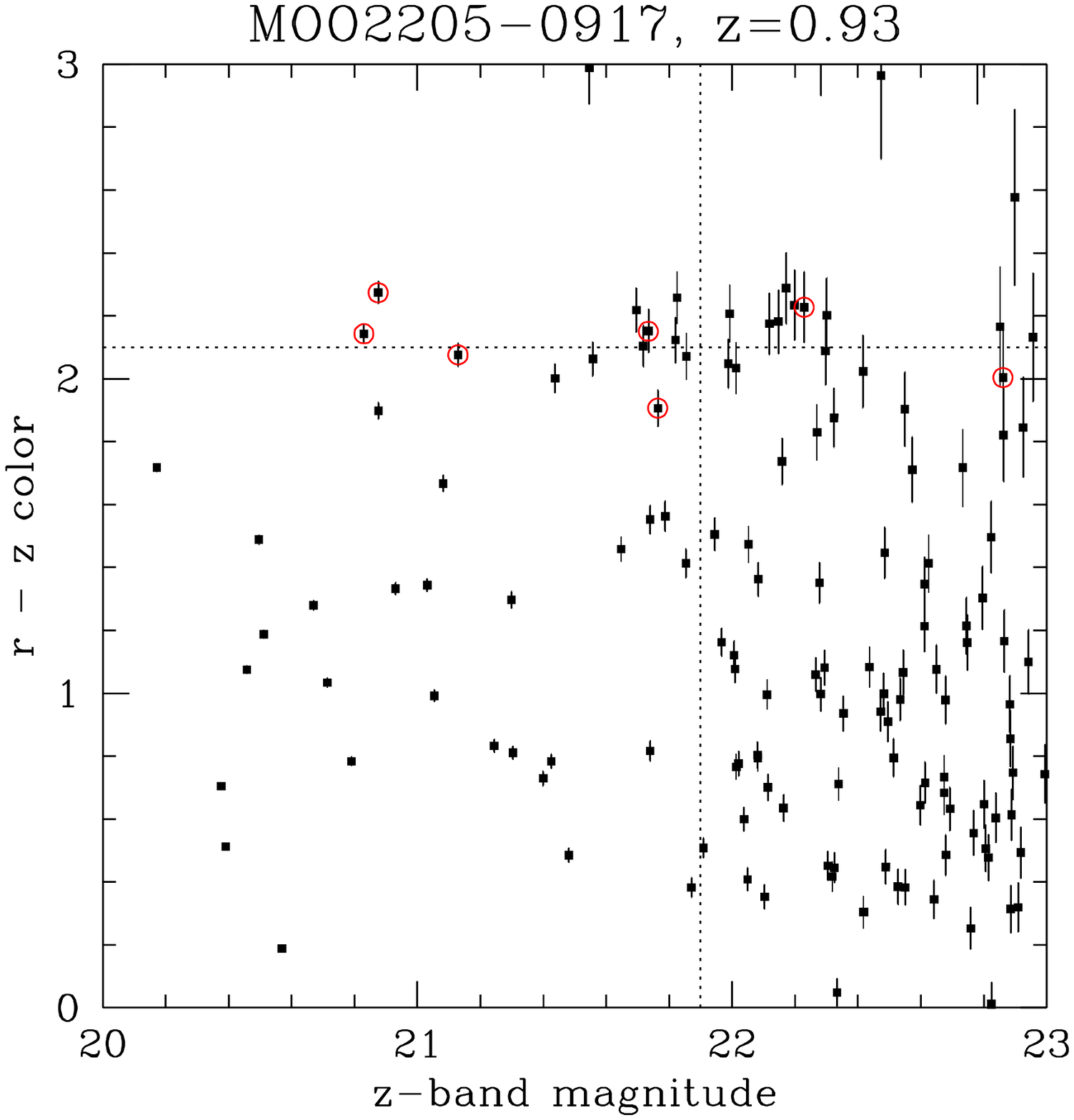}
  }
\fbox{
  \includegraphics[width=0.45\linewidth]{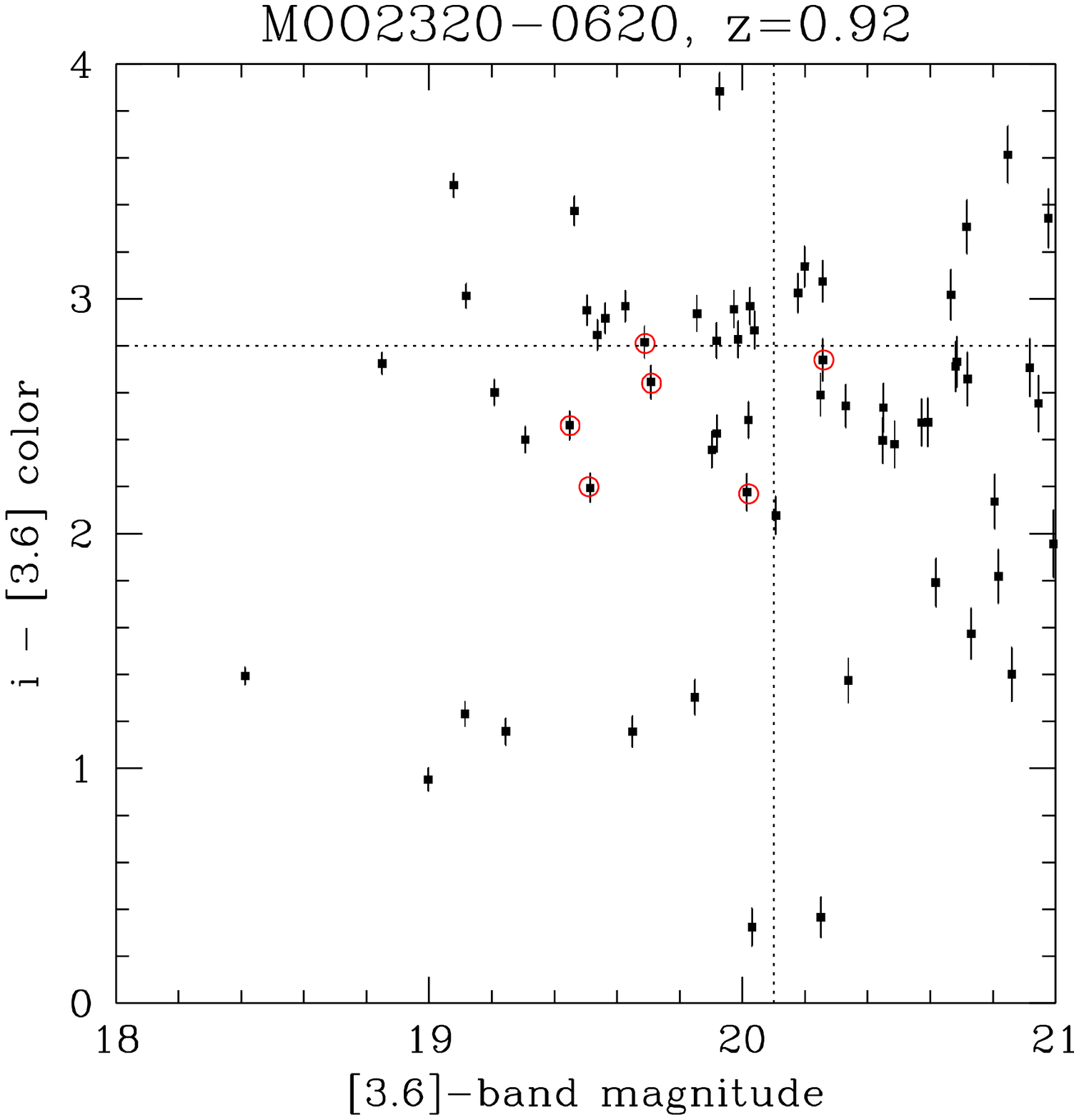}
  }
\fbox{  
    \includegraphics[width=0.45\linewidth]{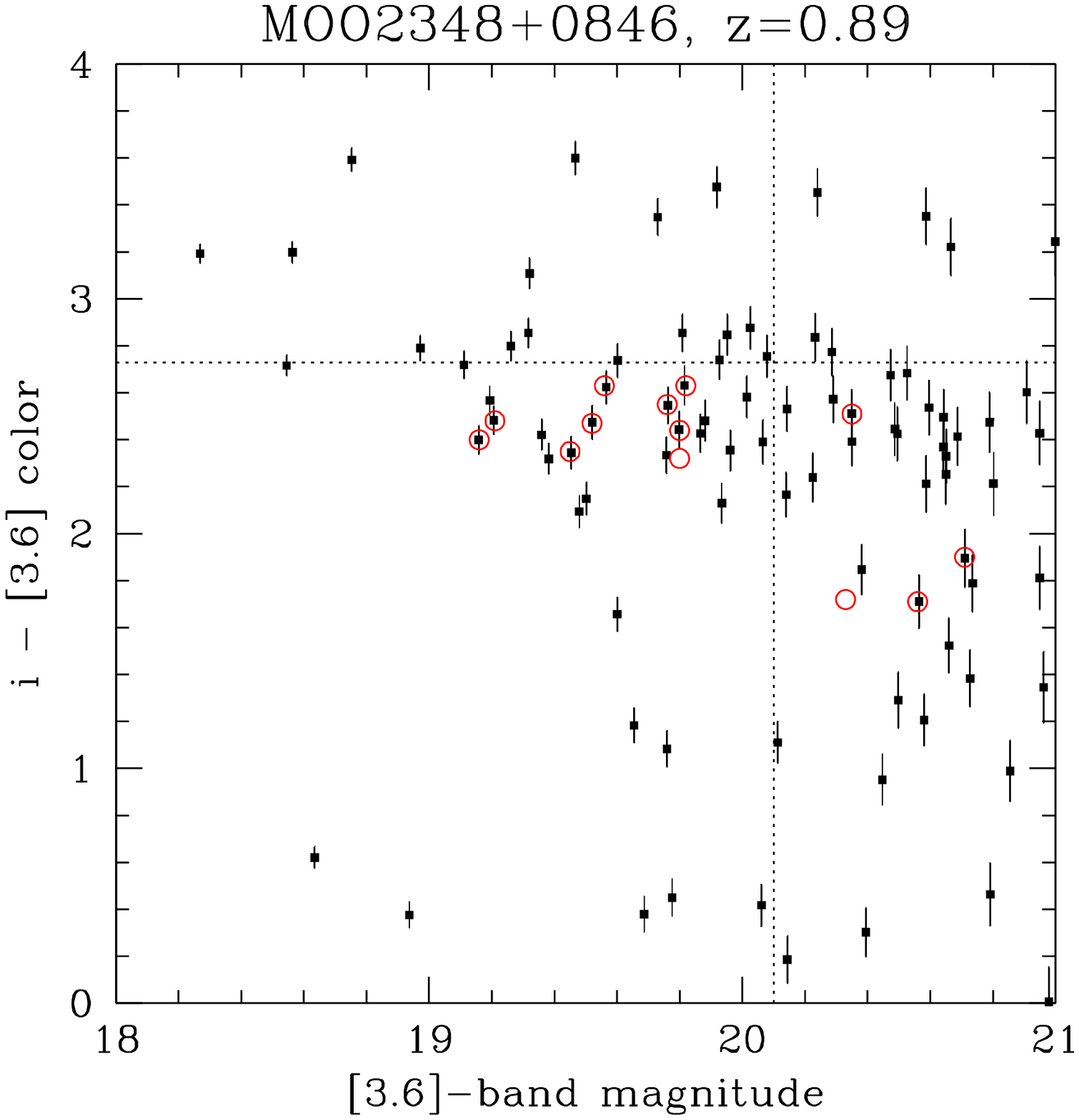}
    }
\centering
\vspace{-1.0cm}
\caption{Continued.}
\end{figure}

\setcounter{figure}{1}

\begin{figure}
   \includegraphics[width=0.45\linewidth]{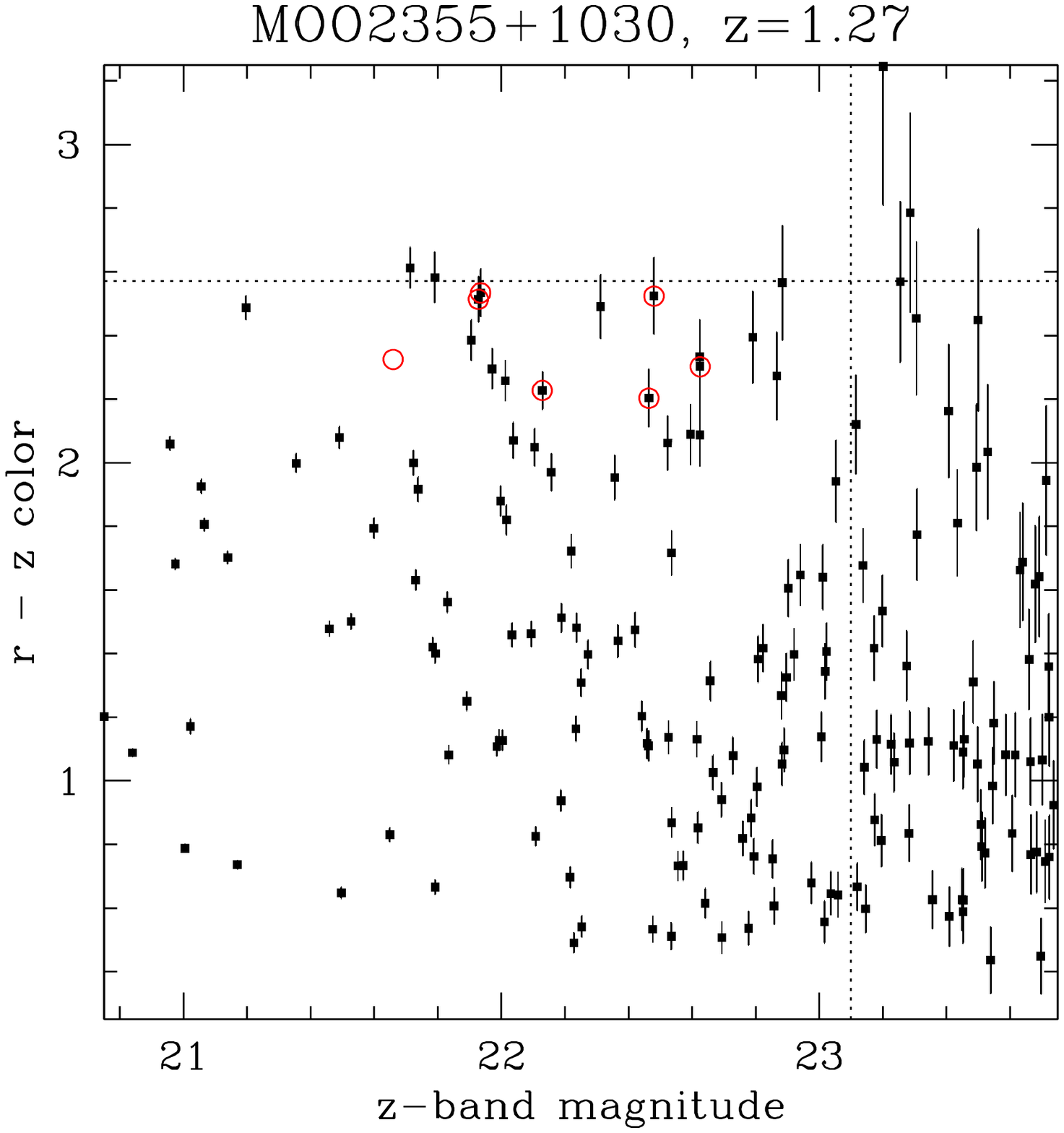}
\caption{Continued.}
\end{figure}

\begin{figure}
\vspace*{-5.0cm}
\centering
\epsscale{0.75}
\plotone{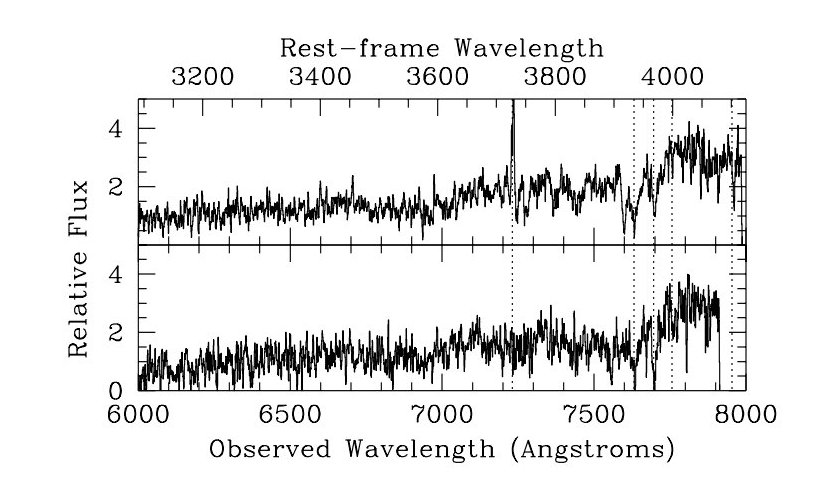}
\plotone{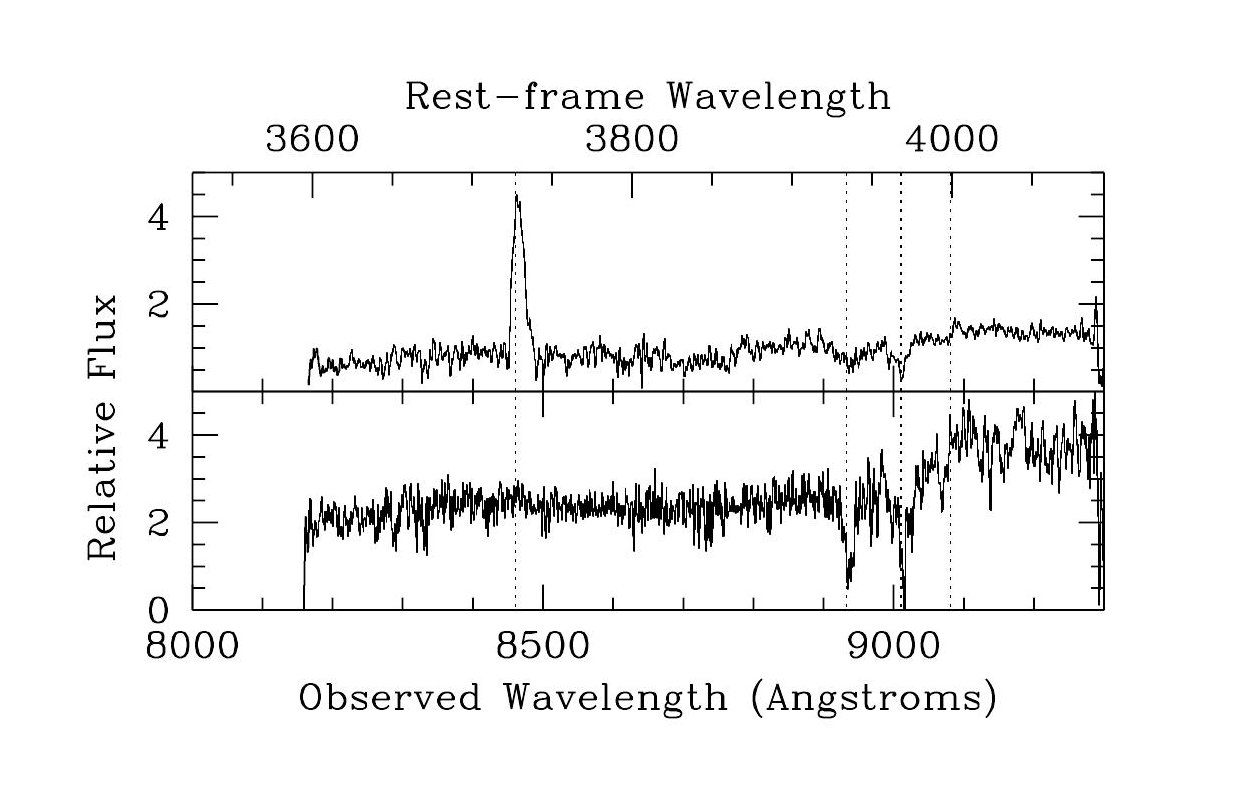}
\caption{Spectra of member galaxies in two clusters, MOO~0012$+$1602 ($z=0.94$) in the top panels and MOO~2355$+$1030 ($z=1.27$) in the bottom panels.   In each pair of panels, the bottom spectrum is of an object whose redshift has a B quality flag, and in the top panel an A quality spectrum. The wavelengths of the following features are shown by the vertical dashed lines: [\ion{O}{2}] $\lambda 3727$, Ca H$+$K, and D4000.  Spectra have been smoothed by a boxcar filter of width 4.5 Angstroms.  }
\label{spec}
\end{figure}

\begin{figure}
\centering
\includegraphics[width=0.45\textwidth]{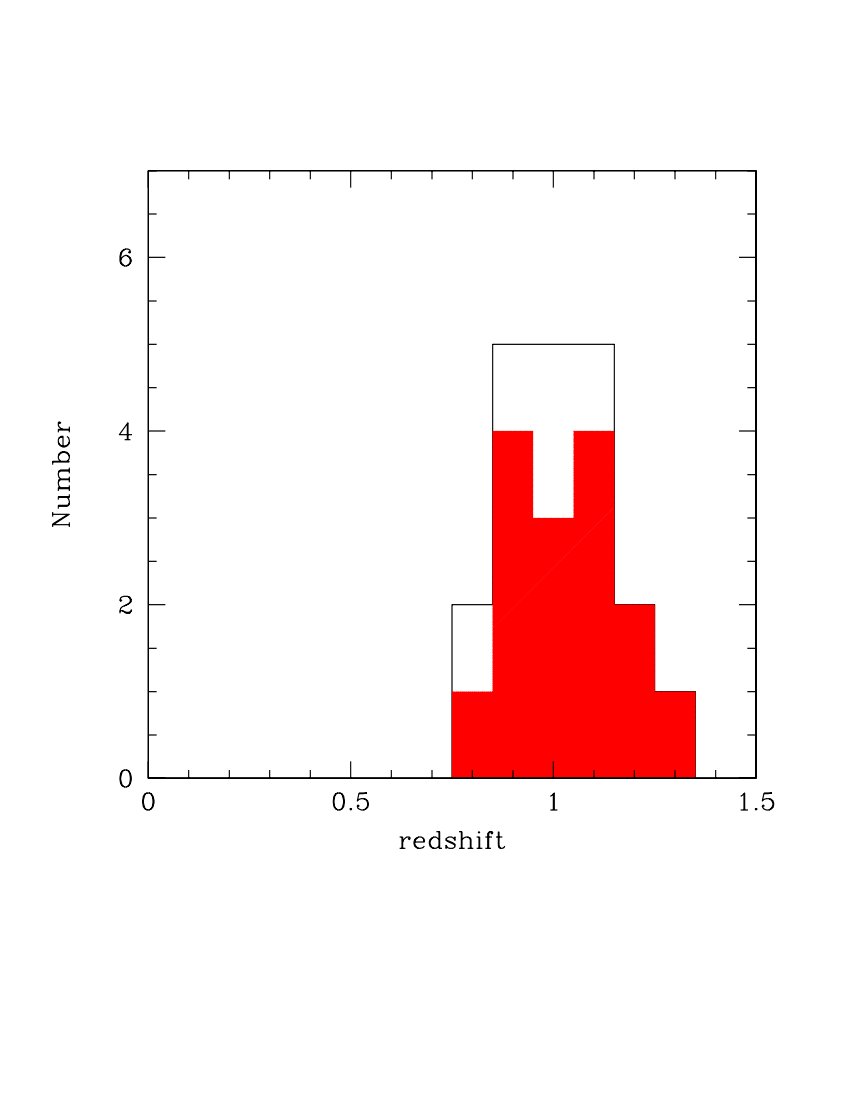}
\caption{Redshift histogram of the clusters confirmed so far in MaDCoWS, including the $z=0.99$ cluster reported in Gettings et al. (2012).  Clusters represented by the open areas were selected in a preliminary search.}
\label{zhist}
\end{figure}

\begin{deluxetable}{lccc}
\tablewidth{0pt}
\tablecaption{MaDCoWS Cluster Redshifts}
\tablehead{ \colhead{Cluster ID} & \colhead{Mean Redshift} & \colhead{Members}  & \colhead{A/E} }
\startdata
MOO~J0012$+$1602\tablenotemark{1}  & 0.944 & 23 & 15/9 \\
MOO~J0024$+$3303 & 1.115 & 6 & 5/1 \\
MOO~J0125$+$1344 & 1.128 & 11 & 6/5 \\
MOO~J0130$+$0922 &  1.146 & 11 & 6/5 \\
MOO~J0133$-$1057 &  0.957 & 13 & 3/10 \\
MOO~J0212$-$1813 &  1.098 & 6 & 4/2 \\
MOO~J0224$-$0620 & 0.816 & 7 & 5/2 \\
MOO~J0245$+$2018 &  0.757 & 14 & 2/12 \\
MOO~J0319$-$0025\tablenotemark{1}   & 1.194 & 20 & 16/4 \\
MOO~J1155$+$3901\tablenotemark{1}  & 1.009 & 7 & 4/3 \\
MOO~J1210$+$3154 & 1.046 & 11 & 7/4 \\
MOO~J1319$+$5519\tablenotemark{1}  & 0.936 & 15 & 11/4 \\
MOO~J1335$+$3004  & 0.980 & 10 & 7/3 \\
MOO~J1514$+$1346\tablenotemark{1}  & 1.059 & 20 & 16/4 \\
MOO~J1625$+$2629 & 1.199 & 7 & 4/3 \\
MOO~J2205$-$0917 & 0.930 & 7 & 7/0 \\
MOO~J2320$-$0620 &  0.923 & 7 & 3/4 \\
MOO~J2348$+$0846 &  0.891 & 16 & 7/9 \\
MOO~J2355$+$1030 & 1.266 & 9 & 5/4 \\
\enddata
\tablenotetext{1}{Sunyaev-Zel'dovich decrements (significance $> 3\sigma$) were measured by CARMA for these clusters (Brodwin et al., in prep.)} 
\tablecomments{The last column provides the numbers of members which have absorption-line (A) and emission-line (E) spectra, as defined in Table 2.}
\label{clusters}
\end{deluxetable}

\end{document}